\documentclass[cernpreprint,atlasdraft=false,UKenglish, texmf, orcidlogo]{atlasdoc}
\usepackage{atlaspackage}
\usepackage{atlasbiblatex}

\usepackage{atlasphysics}
\graphicspath{{logos/}{figures/}}

\usepackage{ANA-STDM-2018-26-PAPER-defs}

\AtlasTitle{Measurement of $W^{\pm}$-boson differential cross-sections in proton--proton collisions with low pile-up data at $\sqrt{s} = 5.02$\,TeV and $13\,\TeV$ with the ATLAS detector}

\AtlasAbstract{%
High precision single-differential $W^\pm$-boson production cross-sections as a function of electron or muon transverse momentum $p_\mathrm{T}$ or their pseudorapity $\eta$, as well as double-differential cross-sections as functions of these variables, are measured in proton--proton collisions at centre-of-mass energies $\sqrt{s}=5.02$\,TeV and 13\,TeV.
The $W$-boson charge asymmetry as a function of lepton $\eta$ is also measured.
The data, collected in dedicated runs at reduced instantaneous luminosity with the ATLAS detector at the Large Hadron Collider, correspond to integrated luminosities of 255\,pb$^{-1}$ at 5.02\,TeV and 338\,pb$^{-1}$ at 13\,TeV. The measurements are in agreement with Standard-Model predictions calculated at next-to-next-to-leading-order in the strong coupling constant $\alpha_s$ including transverse-momentum resummation at next-to-next-to-leading logarithmic accuracy using several  parton distribution functions. The impact of the measured differential cross-sections as a function of lepton $\eta$ on the determination of these functions is studied using a profiling technique.
}

\PreprintIdNumber{CERN-EP-2025-009}

\AtlasJournalRef{\EPJC 85 (2025) 729}
\AtlasDOI{10.1140/epjc/s10052-025-14178-x}

\hypersetup{pdftitle={ATLAS document},pdfauthor={The ATLAS Collaboration}}

\begin{document}

\maketitle

\tableofcontents

%
%

%

\section{Introduction}
\label{sec:intro}

Precise measurements of the differential cross-sections for $W^\pm$-boson production at hadron colliders provide a sensitive probe of perturbative quantum chromodynamics (QCD). Of particular interest is the ability of such measurements to discriminate between different parton distribution functions (PDFs)~\cite{Hou:2019efy,Cridge:2021qfd,Bailey:2020ooq,Cridge:2021pxm,NNPDF:2017mvq,Ball:2021leu,STDM-2020-32}, since the $W$-boson rapidity is strongly correlated with the initial-state parton momentum fraction $x$.
In high-energy proton--proton ($pp$) collisions at the Large Hadron Collider (LHC), the main production mechanism of single $W$ bosons is a quark annihilating with a sea antiquark.
The production at higher $x$ is enhanced by the valence quark contribution which leads to an increase of $W$ bosons boosted in the valence quark direction.
Since the proton contains two valence $u$ quarks and one valence $d$ quark, there is a production asymmetry between $W^+$ and $W^-$ bosons  (referred to as the $W$-boson charge asymmetry), which also varies as a function of rapidity. The boson rapidity cannot be determined unambiguously in leptonic decays of the $W$ boson ($W\to \ell\nu$) because the neutrino passes through the detector unobserved. The charge asymmetry can instead be measured as a function of pseudorapidity ($\eta$) of the charged lepton, which is strongly correlated with the rapidity of the $W$ boson.
In addition, the transverse momentum ($p_\mathrm{T}$) distribution of the charged lepton is sensitive to the mass of the $W$ boson.

High precision predictions with increasingly higher-order corrections are available to compare with data. These predictions incorporate the fixed-order corrections, known up to third order in the strong coupling constant $\alpha_s$~\cite{Duhr:2020sdp,Chen:2022lwc,Campbell:2023lcy}, and the resummation of logarithmic terms from soft and collinear emissions~\cite{Bizon:2019zgf,Neumann:2022lft,Camarda:2023dqn}.

This paper presents a high precision measurement of single-differential cross-sections for $W^\pm$-boson production as a function of lepton $|\eta|$ ($d\sigma/d|\eta|$) and lepton transverse momentum $p_\mathrm{T}$ ($d\sigma/dp_\mathrm{T}$) in the electron and muon ($\ell=e, \mu$) final states.
Double-differential cross-sections as functions of lepton $|\eta|$ and $p_\mathrm{T}$ ($d^2\sigma/d|\eta|dp_\mathrm{T}$) are also measured.
The measurements are based on dedicated data samples collected by the ATLAS experiment at the LHC with low instantaneous luminosity at centre-of-mass energies $\sqrt{s}=5.02$\,TeV and 13\,TeV. These datasets correspond to about 255\,pb$^{-1}$ and 338\,pb$^{-1}$, respectively, with on average two inelastic $pp$ collisions taking place in the same bunch crossing (pile-up).
These low pile-up datasets provide unique and excellent experimental conditions for high precision measurements.
This analysis shares the same datasets, analysis techniques and systematic studies as Ref.~\cite{STDM-2018-17}, where further details of many aspects can be found.
The measurements of the $d\sigma/d|\eta|$, $d\sigma/dp_\mathrm{T}$ and $d^2\sigma/d|\eta|dp_\mathrm{T}$ presented here complement the inclusive cross-sections and single-differential cross-sections as a function of the transverse momentum of the $W^+$, $W^-$ and $Z$ vector bosons ($V=W, Z$) from ATLAS in Ref.~\cite{STDM-2018-17} and the inclusive cross-sections from the CMS experiment~\cite{CMS-SMP-20-004}
using similar data samples. Single-differential cross-sections of $W$-boson production as a function of $\eta$ have also been measured previously by ATLAS at 5.02\,TeV using a much smaller data sample~\cite{HION-2018-02}, 7\,TeV~\cite{STDM-2012-20} and 8\,TeV~\cite{STDM-2017-13}, by CMS at 8\,TeV~\cite{CMS-SMP-14-022}, and by LHCb in the forward region at 7\,TeV and 8\,TeV~\cite{LHCb:2014liz,LHCb:2015mad,LHCb:2016zpq}.
The $W$-boson double-differential cross-section as functions of lepton $\eta$ and $p_\mathrm{T}$ has been measured by CMS at $\sqrt{s} = 13$\,TeV~\cite{CMS-SMP-18-012}.

This paper also presents a measurement of the $W$-boson charge asymmetry ($A_\ell$) as a function of lepton $|\eta|$.
The $W$-boson charge asymmetry has been measured by ATLAS at 5.02\,TeV, 7\,TeV and 8\,TeV~\cite{HION-2018-02,STDM-2012-20,STDM-2017-13}, by CMS at 7\,TeV~\cite{CMS-SMP-12-001,CMS-SMP-12-021}, 8\,TeV~\cite{CMS-SMP-14-022} and $\sqrt{s} = 13$\,TeV~\cite{CMS-SMP-18-012}, and by LHCb at 7\,TeV and 8\,TeV~\cite{LHCb:2014liz,LHCb:2015mad,LHCb:2016zpq}. It has also been measured in proton--antiproton collisions by the CDF and D0 collaborations~\cite{CDF:2009cjw,D0:2013xqc,D0:2014kma,CDF:2021kyj}.

The measurements in the electron and muon channels are combined, and then compared with theoretical predictions at next-to-next-to-leading-order (NNLO) in $\alpha_s$ including transverse-momentum resummation at next-to-next-to-leading logarithmic (NNLL) accuracy with different PDF sets~\cite{Hou:2019efy,Cridge:2021qfd,Bailey:2020ooq,Cridge:2021pxm,NNPDF:2017mvq,Ball:2021leu,STDM-2020-32}. The impact of the measured $d\sigma/d|\eta|$ on the PDFs is further studied using a profiling technique~\cite{Paukkunen:2014zia,HERAFitterdevelopersTeam:2015cre}.


%

\section{ATLAS detector}
\label{sec:detector}

The ATLAS detector~\cite{PERF-2007-01} at the LHC covers nearly the entire solid angle around the collision point.\footnote{ATLAS uses a right-handed coordinate system with its origin at the nominal interaction point (IP) in the centre of the detector and the \(z\)-axis along the beam pipe. The \(x\)-axis points from the IP to the centre of the LHC ring, and the \(y\)-axis points upwards. Polar coordinates \((r,\phi)\) are used in the transverse plane,  \(\phi\) being the azimuthal angle around the \(z\)-axis. The pseudorapidity is defined in terms of the polar angle \(\theta\) as \(\eta = -\ln \tan(\theta/2)\) and is equal to the rapidity
$ y = \frac{1}{2} \ln \left( \frac{E + p_z}{E - p_z} \right) $ in the relativistic limit. Angular distance is measured in units of \(\Delta R \equiv \sqrt{(\Delta y)^{2} + (\Delta\phi)^{2}}\).}
It consists of an inner tracking detector surrounded by a thin superconducting solenoid, electromagnetic and hadronic calorimeters,
and a muon spectrometer incorporating three large superconducting air-core toroidal magnets.

The inner-detector system (ID) is immersed in a \qty{2}{\tesla} axial magnetic field
and provides charged-particle tracking in the range \(|\eta| < 2.5\).
The high-granularity silicon pixel detector covers the vertex region and typically provides four measurements per track,
the first hit generally being in the insertable B-layer (IBL) installed before Run~2~\cite{ATLAS-TDR-19,PIX-2018-001}.
It is followed by the SemiConductor Tracker (SCT), which usually provides eight measurements per track.
These silicon detectors are complemented by the transition radiation tracker (TRT),
which enables radially extended track reconstruction up to \(|\eta| = 2.0\).
The TRT also provides electron identification information
based on the fraction of hits (typically 30 in total) above a higher energy-deposit threshold corresponding to transition radiation.

The calorimeter system covers the pseudorapidity range \(|\eta| < 4.9\).
Within the region \(|\eta|< 3.2\), electromagnetic calorimetry is provided by barrel and
endcap high-granularity lead/liquid-argon (LAr) calorimeters,
with an additional thin LAr presampler covering \(|\eta| < 1.8\)
to correct for energy loss in material upstream of the calorimeters.
Hadronic calorimetry is provided by the steel/scintillator-tile calorimeter,
segmented into three barrel structures within \(|\eta| < 1.7\), and two copper/LAr hadronic endcap calorimeters.
The solid angle coverage is completed with forward copper/LAr and tungsten/LAr calorimeter modules
optimised for electromagnetic and hadronic energy measurements respectively.

The muon spectrometer (MS) comprises separate trigger and
high-precision tracking chambers measuring the deflection of muons in a magnetic field generated by the superconducting air-core toroidal magnets.
The field integral of the toroids ranges between \num{2.0} and \qty{6.0}{\tesla\metre}
across most of the detector.
Three layers of precision chambers, each consisting of layers of monitored drift tubes, cover the region \(|\eta| < 2.7\),
complemented by cathode-strip chambers in the forward region, where the background is highest.
The muon trigger system covers the range \(|\eta| < 2.4\) with resistive-plate chambers in the barrel, and thin-gap chambers in the endcap regions.

The luminosity is measured mainly by the LUCID--2~\cite{LUCID2} detector that records Cherenkov light produced in the quartz windows of photomultipliers located close to the beampipe.

Events are selected by the first-level trigger system implemented in custom hardware,
followed by selections made by algorithms implemented in software in the high-level trigger~\cite{TRIG-2016-01}.
The first-level trigger accepts events from the \qty{40}{\MHz} bunch crossings at a rate close to \qty{100}{\kHz},
which the high-level trigger further reduces in order to record complete events to disk at about \qty{1.25}{\kHz}.

A software suite~\cite{SOFT-2022-02} is used in data simulation, in the reconstruction
and analysis of real and simulated data, in detector operations, and in the trigger and data acquisition
systems of the experiment.


%

\section{Data and simulated event samples}
\label{sec:samples}

This analysis is performed using datasets corresponding to integrated luminosities of $254.9\pm 2.6$\,pb$^{-1}$ at 5.02\,TeV and $338.1\pm 3.1$\,pb$^{-1}$ at 13\,TeV~\cite{DAPR-2021-01}. They were recorded in 2017 and 2018 during dedicated LHC low pile-up runs with an average number of $pp$ interactions of about two, which is about a factor 20 smaller than that of the nominal LHC Run~2 operation between 2015 and 2018. The data was collected with triggers that require at least one muon or electron with transverse momentum thresholds of 14\,GeV and 15\,GeV, respectively~\cite{TRIG-2018-05,TRIG-2018-01,TRIG-2019-03}.

Samples of Monte Carlo (MC) simulated events are used to model the signal and background processes, except multijet production that is modelled from data as discussed in Section~\ref{sec:background}. All MC samples were processed through the full ATLAS detector simulation~\cite{SOFT-2010-01} based on \textsc{Geant4}~\cite{Agostinelli:2002hh} using settings specific to the low pile-up run conditions.
The effects of pile-up collisions in the same or neighbouring bunch crossings were included in the MC simulation by overlaying inelastic $pp$ interactions produced using \textsc{Pythia}\,8.1~\cite{Sjostrand:2007gs} with the NNPDF2.3LO set of PDFs~\cite{Ball:2012cx} and the A3 set of tuned parameters (tune)~\cite{ATL-PHYS-PUB-2016-017}.

The event samples for signal $W$-boson and background $Z$-boson production were generated using the \textsc{Powheg} event generator at next-to-leading-order (NLO) in QCD~\cite{Nason:2004rx,Frixione:2007vw,Alioli:2008gx,Alioli:2010xd} using the CT10 PDF~\cite{Lai:2010vv}, interfaced to \textsc{Pythia}\,8.2~\cite{Sjostrand:2014zea} using the AZNLO tune~\cite{STDM-2012-23}. These \textsc{Powheg+Pythia}\,8 samples were interfaced to \textsc{Photos++}~\cite{Golonka:2005pn} to simulate the effect of final-state QED radiation.
Alternative samples were prepared with \textsc{Sherpa}\,2.2.1 (for 13\,TeV) and 2.2.5 (for 5.02\,TeV)~\cite{Bothmann:2019yzt} using the NNPDF3.0 NNLO PDFs~\cite{Ball:2014uwa} and merging matrix element calculations from \textsc{Comix}~\cite{Gleisberg:2008fv} and \textsc{OpenLoops}~\cite{Buccioni:2019sur,Cascioli:2011va,Denner:2016kdg} for $V+0, 1, 2$ partons at NLO accuracy with $V+3, 4$ partons at leading-order (LO) accuracy in the MEPS@NLO scheme~\cite{Schumann:2007mg,Hoeche:2011fd,Hoeche:2012yf,Catani:2001cc,Hoeche:2009rj}. The $W$-boson signal samples are used to evaluate the uncertainty arising from the choice of MC generator. The $W$ and $Z$-boson samples are normalised to NNLO calculations performed using the \textsc{DYTurbo} program~\cite{Camarda:2019zyx,Camarda:2021ict,Camarda:2021jsw,Camarda:2023dqn}, an optimised version of DYNNLO~\cite{Catani:2007vq,Catani:2009sm}, using the MMHT2014 NNLO PDF set~\cite{Harland-Lang:2014zoa}.
The contribution to the electron and muon final states from leptonically decaying $\tau$-leptons in
$W$-boson decays is treated as background~\cite{ATL-PHYS-PUB-2017-005}.

Background processes from top-antitop-quark pair ($t\bar{t}$) production and single-top-quark production ($Wt$ associated production, $t$-channel, $s$-channel) were generated with \textsc{Powheg+Pythia}\,8~\cite{ATL-PHYS-PUB-2020-023} and normalised to the NNLO predictions with resummation at NNLL accuracy~\cite{Beneke:2011mq,Czakon:2013goa,Czakon:2011xx}. Diboson production $VV$ was generated with \textsc{Sherpa}\,2.2.1 in all decay channels with at least one real lepton in the final state and treated as background~\cite{ATL-PHYS-PUB-2017-005}.


%

\section{Event reconstruction and selection}
\label{sec:selection}

Events are selected if they include exactly one electron or muon candidate that is matched to a corresponding trigger lepton candidate.
Events are also required to have at least one reconstructed collision vertex with two or more charged-particle tracks~\cite{ATL-PHYS-PUB-2015-026}. The vertex with the largest sum of squared transverse momenta of its associated tracks is taken as the primary vertex.

Electron candidates are reconstructed from clusters of energy
deposited in the electromagnetic calorimeter and associated with at
least one track in the ID~\cite{EGAM-2018-01}. Electrons are required to be within the
coverage of the ID and the precision region of the EM calorimeter,
$|\eta| < 2.47$. Electrons in the transition region between the
barrel and endcap calorimeters, $1.37<|\eta|<1.52$, are
excluded. Electron candidates are required to have a transverse
momentum of $p_\mathrm{T} > 25$\,GeV and pass the \textit{Medium} likelihood
identification requirements~\cite{EGAM-2018-01}.  They are also
required to be isolated from nearby activity, as measured by tracks in
a cone of size $\Delta R < 0.2$ around the candidate. The
scalar sum of the $p_\mathrm{T}$ of these tracks ($p_\mathrm{T}^\mathrm{cone20}$) is required to be less than $10\%$ of the electron
$p_\mathrm{T}$ and may not exceed 5\,GeV for electrons with $p_\mathrm{T}  >
50$\,GeV, i.e. $p_\mathrm{T}^\mathrm{cone20}/\min(p_\mathrm{T}, 50\,\mathrm{GeV}) < 0.1$.

The muon reconstruction is performed independently in the ID and in
the MS, and a muon candidate is formed using the combined information from the MS and ID tracks~\cite{MUON-2018-03}. The muon candidates are
required to have an absolute pseudorapidity of $|\eta| < 2.4$, a
transverse momentum of $p_\mathrm{T} > 25$\,GeV and to satisfy the \textit{Medium}
identification criteria~\cite{MUON-2018-03}. Muons are required to be
isolated from nearby activity with the same criterion as electrons.

Lepton candidates are required to originate from the primary
vertex. The track transverse impact parameter significance, $|d_0/\sigma_{d_{0}}|$,
calculated relative to the beam line, must be
smaller than three for muons and smaller than five for
electrons. Furthermore, the longitudinal impact parameter, $z_0$ (the
difference between the $z$-coordinate of the point on the track at
which $d_0$ is defined and the longitudinal position of the primary
vertex), is required to satisfy $|z_0 \sin(\theta)| <$ 0.5 mm.
Dedicated lepton calibrations and efficiency corrections are applied to the reconstructed electron and muon candidates as described in Ref.~\cite{STDM-2018-17}.

The missing transverse momentum, $\vec{p}_\mathrm{T}^\mathrm{miss}$, with its magnitude $E_\mathrm{T}^\mathrm{miss}$, represents a measure of the transverse momentum of the neutrino.
It is defined as $\vec{p}_\mathrm{T}^\mathrm{miss} = - \left(\vec{p}_\mathrm{T} +\vec{u}_\mathrm{T}\right)$ using the lepton $\vec{p}_\mathrm{T}$ and hadronic recoil $\vec{u}_\mathrm{T}$.
The hadronic recoil is reconstructed in the plane transverse to the beam using particle-flow objects  (PFOs), that combine information from charged-particle tracks in the ID and energy deposits in the calorimeter~\cite{PERF-2015-09,JETM-2018-05}.
The hadronic recoil is calibrated using $Z\to \ell^+\ell^-$ events by comparing $u_\mathrm{T}$ with $p_\mathrm{T}^{\ell\ell}$, the transverse momentum of the dilepton system~\cite{STDM-2018-17}.
The resolution of the $u_\text{T}$ is mainly affected by the event activity variable $\Sigma E_\mathrm{T}$, which represents the scalar sum of the transverse energies of all PFOs included in the recoil objects. The modified event activity \setue, defined as $\setue = \Sigma E_\mathrm{T} - u_\text{T}$, primarily reflects contributions from the underlying event and pile-up, and is used to characterise the hadronic recoil resolution. The low pile-up datasets used in this analysis significantly improve the resolution of the recoil measurement.

The background from QCD multijet events is reduced by the requirement $E_\mathrm{T}^\mathrm{miss} > 25$\,GeV. Furthermore, the $W$-boson transverse mass
$m_\mathrm{T} = \sqrt{2 p_\mathrm{T}  E_\mathrm{T}^\mathrm{miss}  (1-\cos\Delta\phi_{\ell\nu})}$ must exceed 50\,GeV, where $\Delta\phi_{\ell\nu}$ is the azimuthal angle
between the lepton $\vec{p}_\mathrm{T}$ and $\vec{p}_\mathrm{T}^\mathrm{miss}$.
After all selections, the numbers of $W\to e\nu$ and $W\to \mu\nu$ candidates for the 5.02\,TeV dataset are $7.1\times 10^{5}$ and $7.5\times 10^{5}$, respectively, and the corresponding number for the 13\,TeV dataset is around $2.2\times 10^{6}$ for both channels.


%

\section{Background estimates and event yields}
\label{sec:background}

Electroweak backgrounds include $W\to \tau \nu$ decays where the $\tau$-lepton decays leptonically, and $Z\to \ell^+\ell^-$ events where one lepton escapes detection. In the electron channel, the $W\to e \nu$ background also includes a contribution from signal $W\to e \nu$ events where the charge of the electron is misreconstructed, forming a background in the oppositely-charged signal sample. In the muon channel, such a charge misreconstruction is negligibly small. There are also background contributions from diboson production and from top quark events that mainly contribute at high lepton $p_\mathrm{T}$.

Background from QCD multijet (MJ) production cannot be reliably
simulated and has to be derived from data~\cite{STDM-2018-17}. Depending on the lepton
flavour, the MJ background has significant contributions from
leptons produced in semileptonic decays of heavy quarks, pion and kaon decays,
or photon conversions.
The multijet yields in the electron and muon channels are estimated separately for positively and negatively charged $W$-boson samples, and also for different centre-of-mass energies.

Four selection regions are used for the multijet background estimate. The signal region (SR) is defined by events satisfying all selection criteria described in Section~\ref{sec:selection}, including $E_\mathrm{T}^\mathrm{miss} > 25$\,GeV and $m_\mathrm{T} > 50$\,GeV. As multijet production is concentrated at lower values of $p_\mathrm{T}$, $E_\mathrm{T}^\mathrm{miss}$ and $m_\mathrm{T}$ as compared to the signal, a fit region (FR) is constructed with the requirements on $E_\mathrm{T}^\mathrm{miss}$ and $m_\mathrm{T}$ dropped, while other selections remain the same as in the SR.
Two control regions, CR1 and CR2, are defined, corresponding to the FR and SR, respectively, except that the lepton in CR1 and CR2 fails the isolation condition.

\begin{figure}[tbp]
\centering
\includegraphics[width=.7\textwidth]{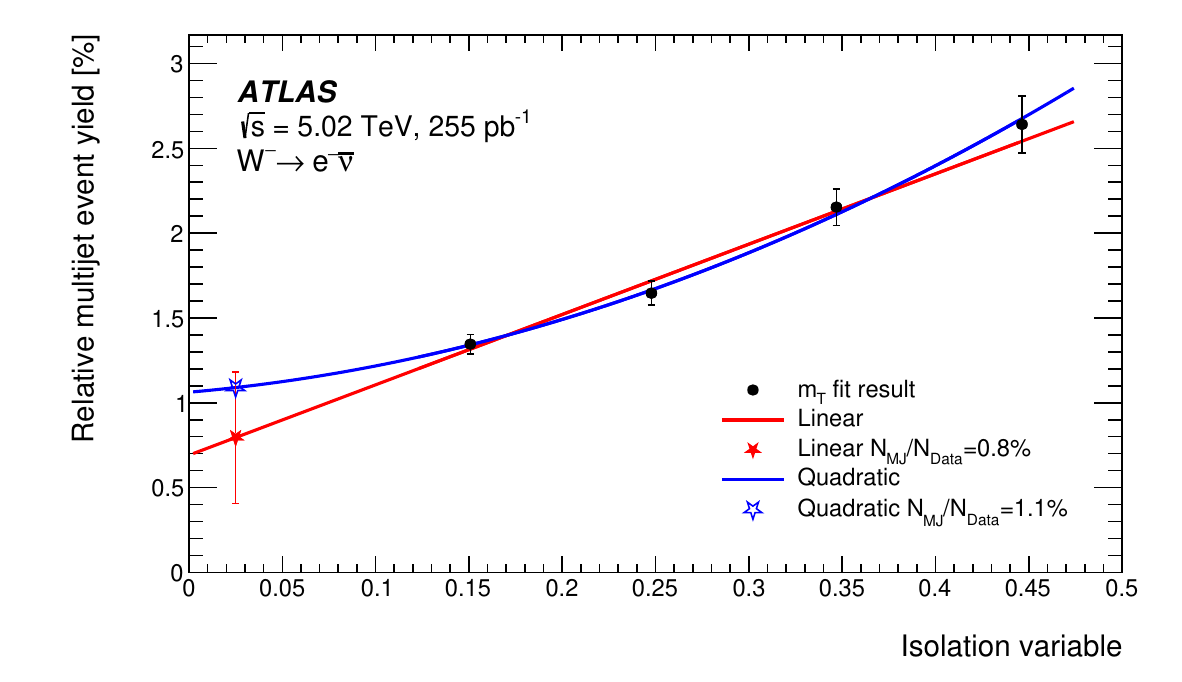}
\caption{Relative multijet event yield in data extrapolated in the signal region from its dependence as a function of isolation variable $p_\mathrm{T}^\mathrm{cone20}/\mathrm{min}(p_\mathrm{T}, 50\,\mathrm{GeV})$ is shown as an example for $W^-\to e^-\bar{\nu}$ at 5.02\,TeV using the $m_\mathrm{T}$ distribution. The vertical error bar on the signal region point at 0.025 represents the total uncertainty including the difference between the linear and quadratic fits.}
\label{fig:multijet}
\end{figure}

The number of multijet events in the FR is derived from a template fit to the data distributions as a function of $p_\mathrm{T}$, $E_\mathrm{T}^\mathrm{miss}$ or $m_\mathrm{T}$, exploiting the different shapes of multijet events, $W\to\ell\nu$ signal, electroweak and top-quark background components in these variables.
The multijet template used in the FR is derived from the data distributions in the CR1, after subtracting the small $W\to\ell\nu$ signal and non-multijet background components using simulation.
The other components in the FR are taken from simulation.
In a final step, the multijet yields in the SR are determined by multiplying the FR yields by a transfer factor that corrects for the acceptance of the $E_\mathrm{T}^\mathrm{miss}$ and $m_\mathrm{T}$ selections and the dependence on the isolation requirement. The transfer factor is determined by calculating the ratio of multijet event yields in the CR2 and the CR1.
As the transfer factor depends on the isolation criterion, mutually exclusive intervals
in the isolation variable are chosen to create statistically
independent samples
that are progressively closer to the signal-candidate selection.
Figure~\ref{fig:multijet} shows the dependence on the isolation variable interval of multijet yields derived using these transfer factors, taking the $m_\mathrm{T}$ variable as an example.
A linear extrapolation to the signal region is performed, with the difference in normalisation with respect to a quadratic extrapolation included as an additional systematic uncertainty. The other dominant multijet yield uncertainties include the statistical uncertainty from the finite data and MC sample size in the fit regions, and those arising from the linear extrapolation to the average isolation value in the signal region and the possible mismodelling of the jet activity in those regions failing the isolation requirement~\cite{STDM-2018-17}.
In addition, a shape uncertainty is obtained by taking the difference between the extrapolated shape in the SR and the multijet shape in the isolation interval $[0.1, 0.2]$ in the CR2.
The multijet yields are estimated by using template fits performed as functions of $p_\mathrm{T}$,  $E_\mathrm{T}^\mathrm{miss}$ and $m_\mathrm{T}$ separately. The results are found to be consistent and their average is used for the final result.

\begin{figure}[tbp]
\centering
\subfloat[]{{\includegraphics[width=0.495\linewidth]{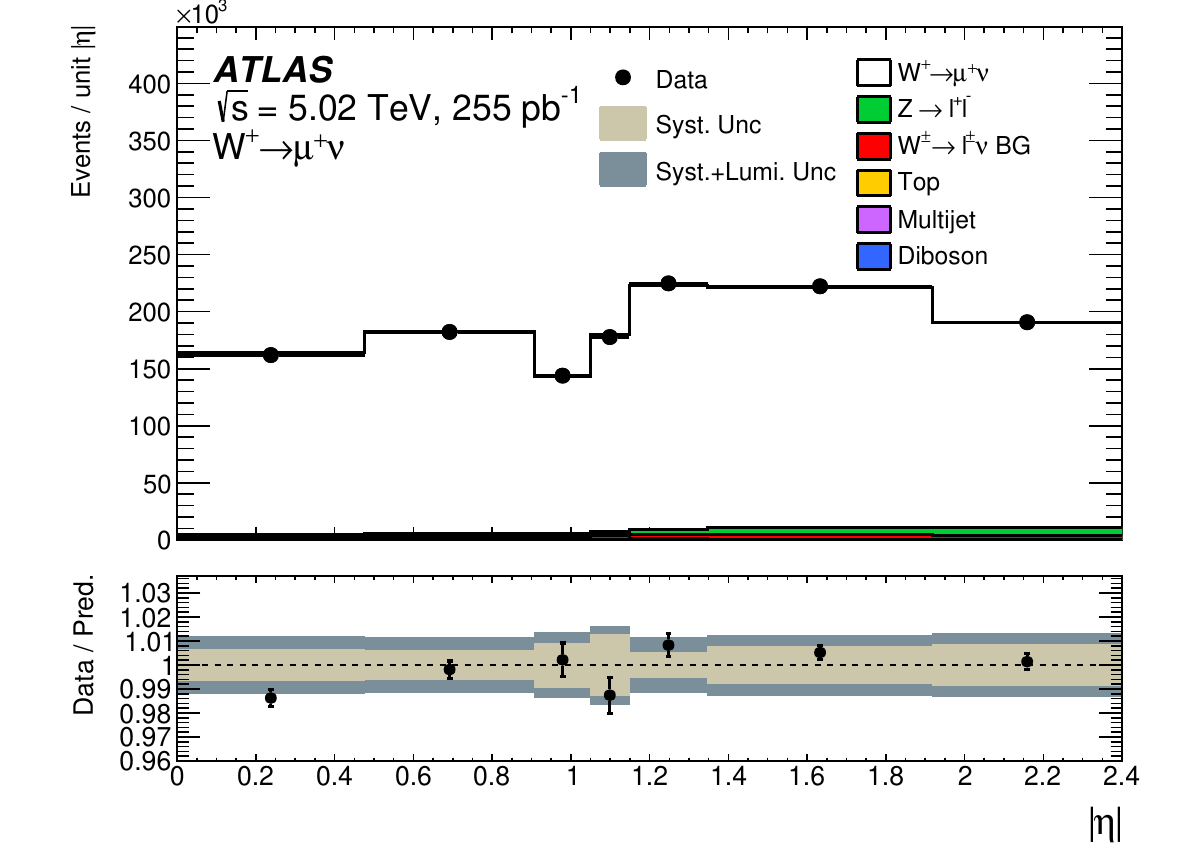} }}%
\subfloat[]{{\includegraphics[width=0.495\linewidth]{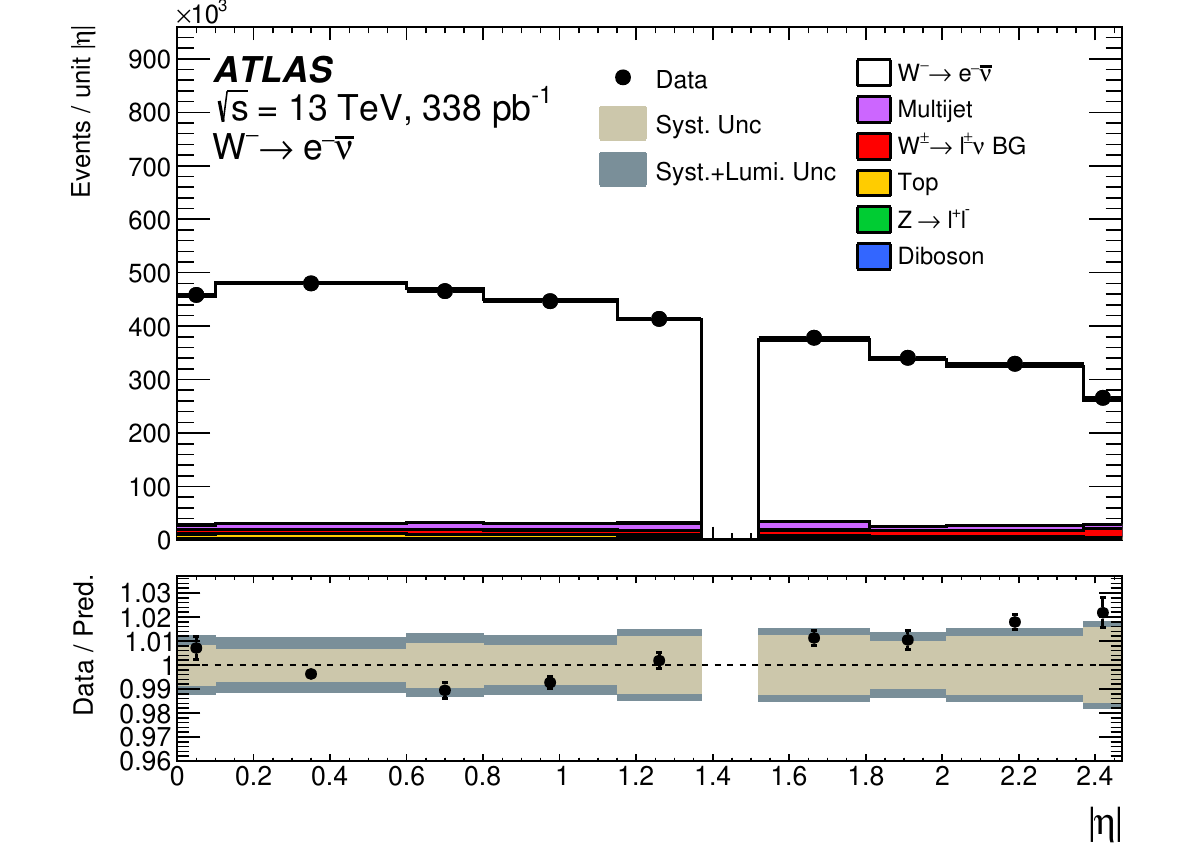} }}\\
\subfloat[]{{\includegraphics[width=0.495\linewidth]{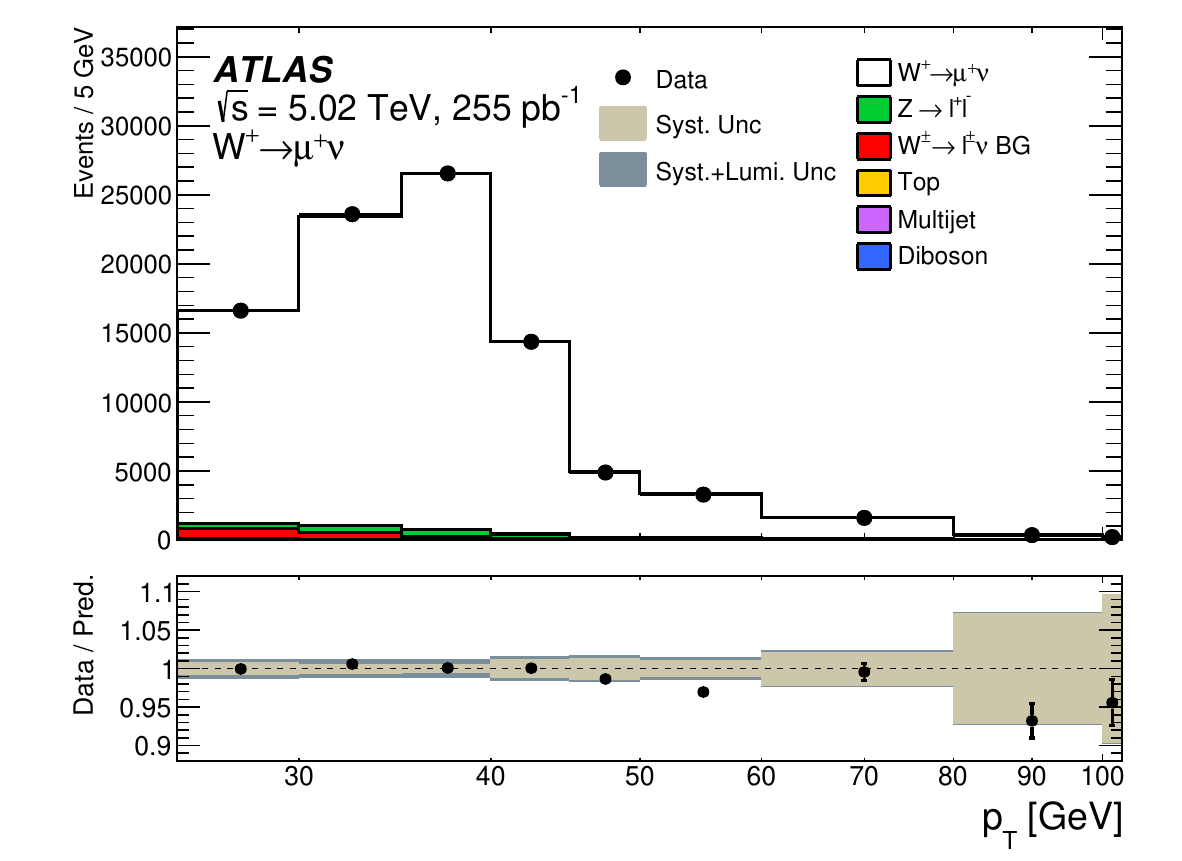} }}
\subfloat[]{{\includegraphics[width=0.495\linewidth]{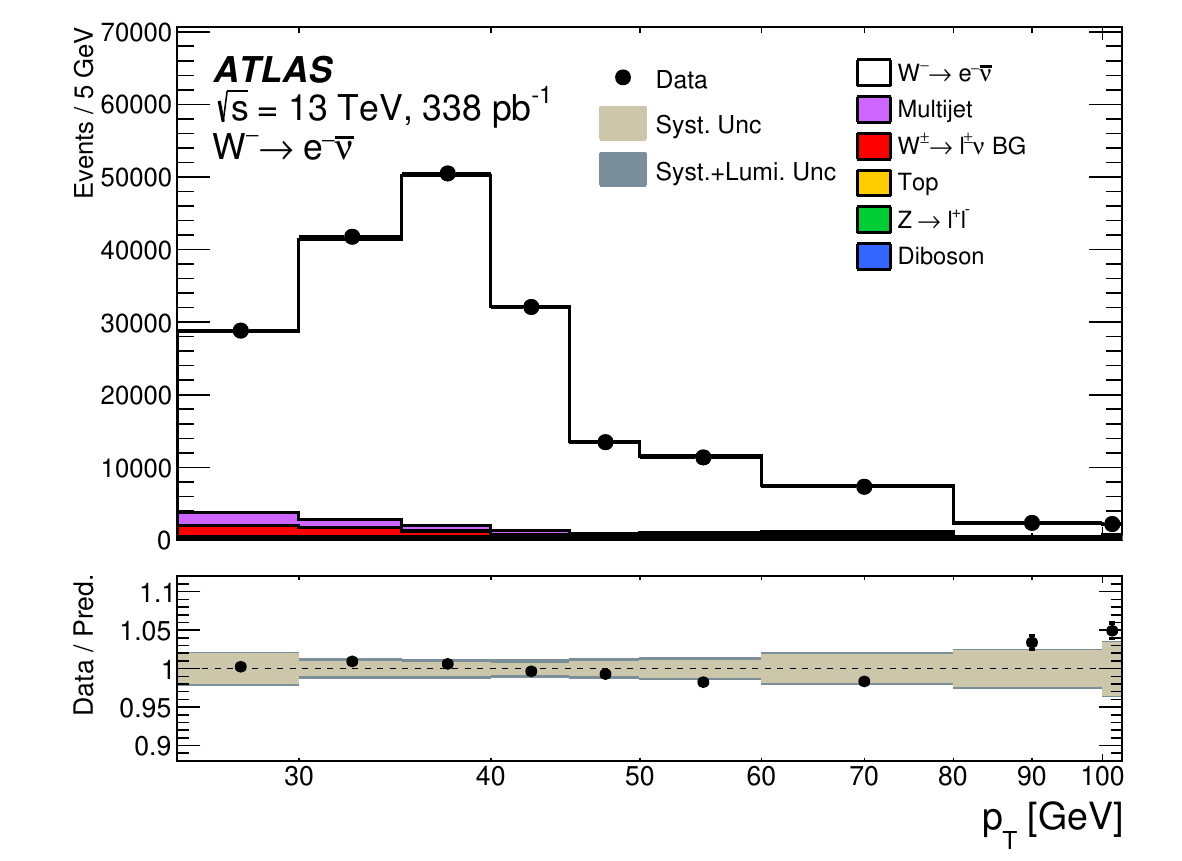} }}\\
\caption{Detector-level distributions of lepton $|\eta|$ in the (a) 5.02\,TeV $W^+\to \mu^+\nu$ and (b) 13\,TeV $W^-\to e^-\bar{\nu}$ channels for data, and expected signal and various background contributions normalised to the integrated luminosity of the data sample. Similar detector-level distributions of lepton $p_\mathrm{T}$ in the (c) 5.02\,TeV $W^+\to \mu^+\nu$ and (d) 13\,TeV $W^-\to e^-\bar{\nu}$ channels. The last $p_\mathrm{T}$ bin has a width of 5\,GeV but includes all events with $p_\mathrm{T}>100$\,GeV. The bottom panel displays the ratio of data over expectation, with the data statistical error shown by the error bar on the points. The dark shaded band represents the total systematic uncertainty, while the light shaded band shows the total systematic uncertainty excluding the luminosity uncertainty.}
\label{fig:kine1d}
\end{figure}

\begin{figure}[tbp]
\centering
\subfloat[]{{\includegraphics[width=0.95\linewidth]{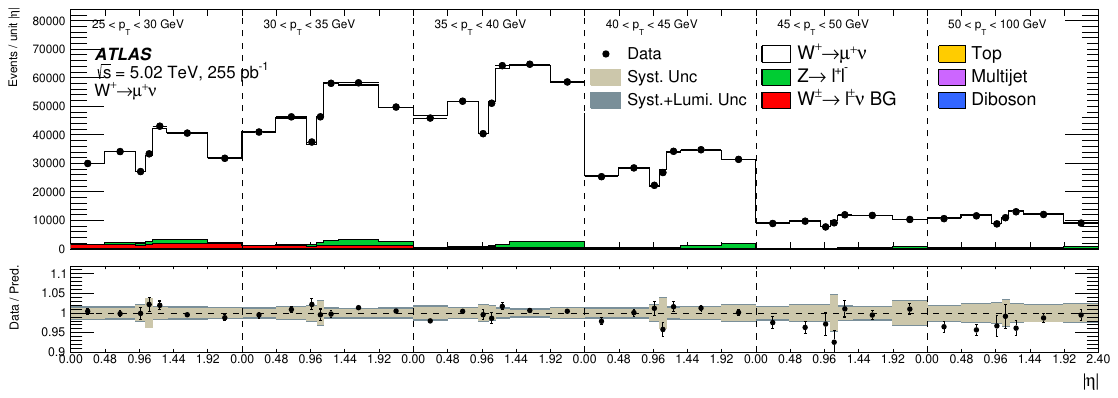} }}\\
\subfloat[]{{\includegraphics[width=0.95\linewidth]{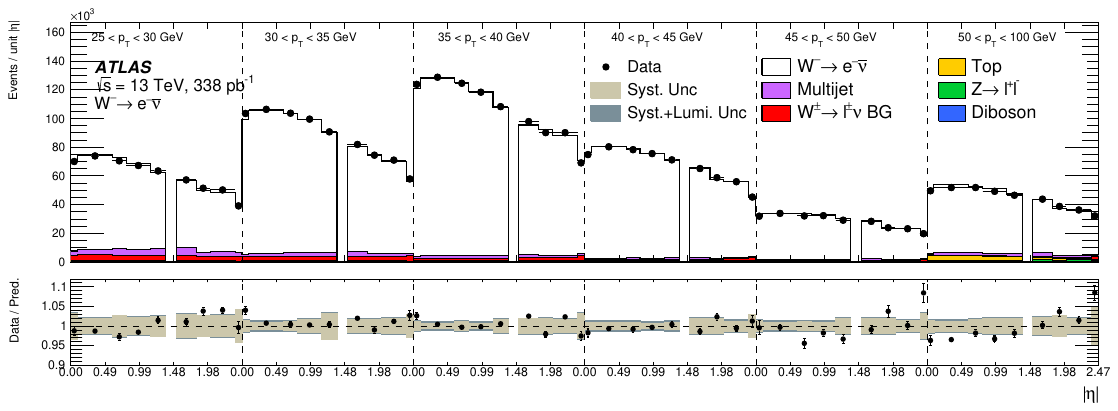} }}
\caption{Detector-level distributions of lepton $|\eta|$ for various ranges of lepton $p_\mathrm{T}$ in the (a) 5.02\,TeV $W^+\to \mu^+\nu$ and (b) 13\,TeV $W^-\to e^-\bar{\nu}$ channels for data, and expected signal and various background contributions normalised to the integrated luminosity of the data sample. The bottom panel displays the ratio of data over expectation, with the data statistical error shown by the error bar on the points. The dark shaded band represents the total systematic uncertainty, while the light shaded band shows the total systematic uncertainty excluding the luminosity uncertainty.}
\label{fig:kine2d}
\end{figure}

The expected event yields for signal and background processes are shown in Table~\ref{tab:event-yields}.
The total background level ranges between 3\% and 7\% depending on the channel and the centre-of-mass energy.
Figure~\ref{fig:kine1d} shows the detector-level $|\eta|$ and $p_\mathrm{T}$ distributions for two  selected channels at 5.02\,TeV and 13\,TeV. The two-dimensional distributions of $\eta$ and $p_\mathrm{T}$ are presented as consecutive distributions of \AbsEta\ for different ranges of \pTl, as shown in Figure~\ref{fig:kine2d}.
Overall, good agreement is observed between the data and the expectations in all the kinematic distributions.
The choice of the bin boundaries for the \AbsEta\ distribution is different for the electron and muon channels, driven by details of the detector geometry. The bin boundaries for the $p_\mathrm{T}$ distribution are the same for electrons and muons, and correspond to those for the unfolded fiducial cross-section as discussed in Section~\ref{sec:procedure}. The hole in the $|\eta|$ distributions in the electron channel corresponds to the excluded barrel-endcap calorimeter transition region ($1.37<|\eta|<1.52$) in the selection. The differences between the electron and muon distributions reflect both the different efficiencies for reconstructing and selecting the two lepton flavours, and the intrinsic kinematic differences between $W^+$-boson events at 5.02\,TeV and $W^-$-boson events at 13\,TeV.

\begin{table}[tb]
\caption{Observed and expected event yields for the
$W$-boson selections in the 5.02\,TeV and 13\,TeV datasets
for the electron and muon channels split by the charge of the
reconstructed lepton.
Fractions in percent for the signal and background processes relative to the total expected event yield are also shown, where the uncertainties for the
multijet background correspond to the total uncertainties.
The uncertainty in the expected event yield includes the simulation statistical uncertainty and that of the multijet background.
The category $W\to\ell\nu$ BG refers to background from $W\to\tau\nu$ as well
as charge misidentification in the electron channels.}
\label{tab:event-yields}
\setlength\tabcolsep{3pt}
\resizebox{\textwidth}{!}{

\begin{tabular}{l|r|r||c|c|c|c|c|c}
\toprule
Channel &  Observed  &  \multicolumn{1}{c||}{Expected}  & Signal [\%] &  {$W\to\ell\nu$ BG}  [\%] &  {$Z \rightarrow \ell\ell$} [\%] &  {Top} [\%] &  {Diboson} [\%] &  {Multijet} [\%] \\
\midrule
\multicolumn{9}{c}{$\sqrt{s} = 5.02$\,TeV \; $W\to \ell\nu$}\\
\midrule
$W^- \rightarrow e^-\bar{\nu}$   & 274375 & 276000  $\pm$  1000 & 95.8  & 2.6  & 0.4  & 0.3 & 0.1 & 0.8  $\pm$  0.4 \\
$W^+ \rightarrow e^+\nu$  & 430662 &431000 $\pm$ 1000 & 96.8  & 2.1 & 0.3 & 0.2 & 0.1 & 0.5 $\pm$ 0.2 \\
$W^- \rightarrow \mu^-\bar{\nu}$   & 288026 & 289000 $\pm$ 1000 & 94.8 & 1.8 & 2.9 & 0.3 & 0.1 & 0.1 $\pm$ 0.3 \\
$W^+ \rightarrow \mu^+\nu$  & 457223  & 457000 $\pm$ 1000 & 95.8 & 1.7 & 2.1 & 0.2 & 0.1 & 0.1 $\pm$ 0.2 \\
\midrule
\multicolumn{9}{c}{$\sqrt{s} = 13$\,TeV \; $W\to\ell\nu$}\\
\midrule
$W^- \rightarrow e^-\bar{\nu}$ & 949297 & 947000 $\pm$ 6000 & 92.6 & 2.4 & 0.7 & 1.3 & 0.2 & 2.9 $\pm$ 0.6 \\
$W^+ \rightarrow e^+\nu$  & 1207652 & 1192000 $\pm$ 7000 & 93.7 & 2.1 & 0.6 & 1.1 & 0.1 & 2.4 $\pm$ 0.6 \\
$W^- \rightarrow \mu^-\bar{\nu}$ & 964514 & 966000 $\pm$ 4000 & 92.9 & 1.5 & 3.5 & 1.2 & 0.2 & 0.6 $\pm$ 0.4 \\
$W^+ \rightarrow \mu^+\nu$  & 1245755 & 1230000 $\pm$ 4000 & 93.8 & 1.5 & 3.1 & 1.0 & 0.1 & 0.5 $\pm$ 0.3 \\
\bottomrule
\end{tabular}


}
\end{table}

%

\section{Measurement procedure}
\label{sec:procedure}

The differential cross-sections at the particle level are measured in the same fiducial phase space as in Ref.~\cite{STDM-2018-17}, corresponding to Born-level leptons with $\pTl>25$\,GeV, $\AbsEta<2.5$, $\mT>50$\,GeV and $p_\mathrm{T}^\nu>25$\,GeV. The bins defined by particle-level kinematics for the single-differential cross-sections in lepton $|\eta|$ and $p_\mathrm{T}$ are:
\begin{itemize}
\item $|\eta|$: $[0, 0.42, 0.84, 1.05, 1.37, 1.52, 1.95, 2.5]$,
\item $p_\mathrm{T}$: $[25, 30, 35, 40, 45, 50, 60, 80, 100]$\,GeV.
\end{itemize}
The bins are chosen by considering the sensitivity to the underlying physics, detector resolution effects, detector geometry, and the statistical precision in each bin. To simplify the combination process described below, a common binning scheme is chosen for the electron and muon channels.
For the double-differential cross-section measurement, the \AbsEta\ bin boundaries from the single-differential measurement are used, while the last three $p_\mathrm{T}$ bins from $50-100$\,GeV are merged into a single bin due to the small number of events with high $p_\mathrm{T}$.

To derive the differential cross-sections, the distributions observed at the detector level after analysis selection and background subtraction are corrected for detector effects using an iterative Bayesian unfolding method~\cite{DAgostini:1994fjx,DAgostini:2010hil}. First, the data are corrected for events that pass the detector-level selection but not the particle-level selection, with purity correction factors derived from simulated samples. The iterative Bayesian unfolding technique is then used to correct for the finite detector acceptance, resolution and reconstruction efficiency in order to estimate the true underlying distribution. Simulated events are used to determine the response matrices needed to correct for the migration between bins in the detector-level distributions. The unfolding procedure follows Ref.~\cite{STDM-2018-17} and determines the best estimate of the underlying distribution using:
\begin{equation}
\Tilde{u}_j = \sum_i U_{ji}  (d_i - b_i)p_{i}\,,
\end{equation}
where the indices $i$ and $j$ refer to bin $i$ and $j$ in the detector-level and unfolded or true distributions, respectively, $\Tilde{u}_j$ is the unfolded spectrum,
and $U$ symbolically represents the unfolding transformation, which in the simplest case is just the inverse of the response matrix. The background-subtracted data distribution $d_i - b_i$ is multiplied by the purity correction factors $p_{i}$. The purity correction factors $p_{i}$ are defined by the number of generated and reconstructed events in bin $i$ divided by the number of reconstructed events in the bin. The number of iterations in the unfolding procedure is optimised to minimise the total measurement uncertainty, and specifically the uncertainty related to possible biases induced by the unfolding. Finally, the differential cross-section in bin $j$ is derived by dividing $\Tilde{u}_j$ by the product of the luminosity, bin width and efficiency correction factors. The latter are defined by the number of generated and reconstructed events in bin $j$ divided by the number of generated events in the bin.

The $W$-boson production differential cross-sections are measured separately in the electron and muon channels. The measured cross-sections in the two channels are compared and are found to be consistent within the uncorrelated uncertainties. They are then combined as in Ref.~\cite{STDM-2018-17} by using an iterative implementation~\cite{Dembinski:2018ihc} of the Best Linear Unbiased Estimator (BLUE) method~\cite{Lyons:1988rp, Valassi:2003mu}. The iterative procedure reduces the bias resulting from non-Gaussian uncertainty sources.

The $W$-boson charge asymmetry $A_\ell$ is extracted using the formula:
\begin{equation}
A_{\ell}(|\eta|) = \displaystyle\frac{\frac{d\sigma_{W^{+}}}{d\AbsEta} - \frac{d\sigma_{W^{-}}}{d\AbsEta}}{\frac{d\sigma_{W^{+}}}{d\AbsEta} + \frac{d\sigma_{W^{-}}}{d\AbsEta}}\,,\label{eq:asymmetry}
\end{equation}
where the differential cross-sections as a function of lepton $|\eta|$ are obtained from the combination of the electron and muon channels.


%

\section{Systematic uncertainties}%
\label{sec:Systematic uncertainties}

The uncertainties in the integrated luminosities are $1.0\%$ and
$0.92\%$ for the 5.02\,TeV and 13\,TeV datasets, respectively, following the methodology
discussed in Ref.~\cite{DAPR-2021-01}, using the LUCID-2
detector~\cite{LUCID2} for the primary luminosity measurements,
complemented by measurements using the inner detector and
calorimeters. These are the leading systematic uncertainties for the single-differential cross-section measurements except at the high lepton $p_\mathrm{T}$ tails.

The electron efficiency corrections and corresponding uncertainties as functions of lepton $p_\mathrm{T}$ and $|\eta|$ are determined following the strategies outlined in Ref.~\cite{EGAM-2018-01}. The reconstruction efficiency is extrapolated from corrections derived using datasets with high pile-up conditions~\cite{STDM-2018-17}. The remaining efficiency corrections and their corresponding uncertainties are obtained using $Z\to e^+e^-$ events in the low pile-up datasets.
The electron energy calibration uncertainty is also determined using the low pile-up datasets and the methods described in Ref.~\cite{EGAM-2018-01}. These uncertainties are typically below 0.3\% for the single-differential cross-section measurements except at the low and high lepton $p_\mathrm{T}$ regions, where they reach up to 1\%.
The charge misidentification probability and the corresponding uncertainties are evaluated as a function of lepton $\eta$. The impact of charge misidentification is very small for the cross-section measurement but contributes up to 0.4\% when determining the charge asymmetry.

The muon efficiency corrections and corresponding uncertainties follow a similar approach to those for electrons. Most of the efficiencies are measured using $Z\to \mu^+\mu^-$ events in the low pile-up dataset following Ref.~\cite{MUON-2018-03}, while the reconstruction efficiency is extrapolated from datasets with high pile-up conditions~\cite{STDM-2018-17}.
The muon energy calibration uncertainty is derived using the Run~2 dataset with high pile-up conditions~\cite{MUON-2022-01} and then extrapolated to the low pile-up conditions. These uncertainties are typically at up to a few permille level.

The uncertainties arising from the hadronic recoil calibration are grouped into several categories~\cite{STDM-2018-17}. The primary uncertainty arises from the corrections derived from $Z$-boson events.
The corrections to the simulation to account for mismodelling of the data \setue\ distribution give rise to subdominant uncertainty sources in the single-differential lepton $p_\mathrm{T}$ cross-section measurements in the high lepton $p_\mathrm{T}$ regions.

Uncertainties due to the unfolding procedure are determined following Ref.~\cite{STDM-2016-03}, based on the regularisation and differences in shape between the data and MC simulations. The particle-level spectra in simulation are reweighted using a smooth function so that the reconstructed-level spectra match the data distributions. The reweighted MC spectra are then treated as pseudodata and unfolded back to the particle-level. The differences between the unfolded and reweighted particle-level spectra are considered as the unfolding bias, which has only a negligible effect on the results.

The uncertainties originating from the finite size of the MC simulation samples are treated as systematic uncertainties, and they are estimated by using the bootstrap method~\cite{ATL-PHYS-PUB-2021-011}. They only have a sizeable impact in the highest lepton $p_\mathrm{T}$ bin of the single-differential lepton $p_\mathrm{T}$ distributions. They have also a contribution around 0.5\% for the $W$-boson charge asymmetry measurement.

The uncertainties arising from the choice of the generator for the \W\ signal samples are estimated by using alternative \textsc{Sherpa} \W\ samples, which are reweighted to have the same PDFs and calibrated to the default \textsc{Powheg}+\textsc{Pythia} \W\ samples for modelling the hadronic recoil and primary vertex reconstruction efficiency~\cite{STDM-2018-17}.
They are the dominant or next-most-dominant systematic uncertainty affecting the cross-section results by up to 5\% at the highest $p_\mathrm{T}$ bin at 5.02\,TeV and the charge asymmetry results by up to 1.5\%.

The uncertainties related to the background modelling of the electroweak and top-quark background processes are estimated from simulated samples by varying the PDF input, $\alpha_\mathrm{s}$, and QCD renormalisation and factorisation scales to cover for missing higher-order contributions. They are 5\% for the electroweak backgrounds involving $W$ and $Z$ bosons, 7\% for $t\bar{t}$ processes, and 10\% for diboson and single-top processes. They have negligible impact on the cross-section measurements given the small background contributions.
On the other hand, the multijet background uncertainty is dominant at 13\,TeV in particular in the low lepton $\pTl$ region.

The combination of the electron and muon decay channel results, and the determination of the $W$-boson charge asymmetry, require a model for the correlation of systematic uncertainties between decay channels and $W$-boson charges. The PDF profiling study described in Section~\ref{sub:PDF profiling} also requires the correlations between centre-of-mass energies.
The following correlation model of the uncertainties is applied across the various lepton decay channels, $W$-boson charges and centre-of-mass energies~\cite{STDM-2018-17}.
The lepton efficiency corrections, trigger efficiency, and energy calibrations are considered fully correlated for the measurements at the same centre-of-mass energy, but uncorrelated between different energies. The latter choice has, however, a negligible impact on the results related to the PDF study.
The recoil calibration uncertainty, determined using $Z$-boson events, is treated as fully correlated across the $W^+$ and $W^-$ measurements and between lepton flavours. However, the recoil calibration remains independent between the 5.02\,TeV and 13\,TeV datasets. The subdominant component, which accounts for discrepancies between data and MC in $\sum\bar{E}_\mathrm{T}$ distributions, is treated as uncorrelated.
Uncertainties in electroweak and top-quark backgrounds, are assumed to be fully correlated across all measurements.
For the multijet background uncertainty, the statistical component is considered uncorrelated between different lepton decay channels, $W$-boson charges and beam energies, while methodological variations are assumed to be fully correlated among all measurements.
The unfolding bias uncertainty is treated as uncorrelated between different measurements. Generator uncertainties are fully correlated between all measurements.


%

\section{Results}%
\label{sec:Results}

In this section, the single- and double-differential cross-section results are presented first, followed by the $W$-boson charge asymmetry results. The impact of the new single-differential cross-sections as a function of lepton $|\eta|$ on PDFs is then presented. When integrating the differential cross-sections over all the full fiducial phase space defined in Section~\ref{sec:procedure}, the inclusive cross-section values are compatible with those reported in Ref.~\cite{STDM-2018-17}.

\subsection{Differential cross-sections}%
\label{sub:cross-sections}

\begin{figure}[tbp]
\centering
\subfloat[]{\includegraphics[width=0.495\linewidth]{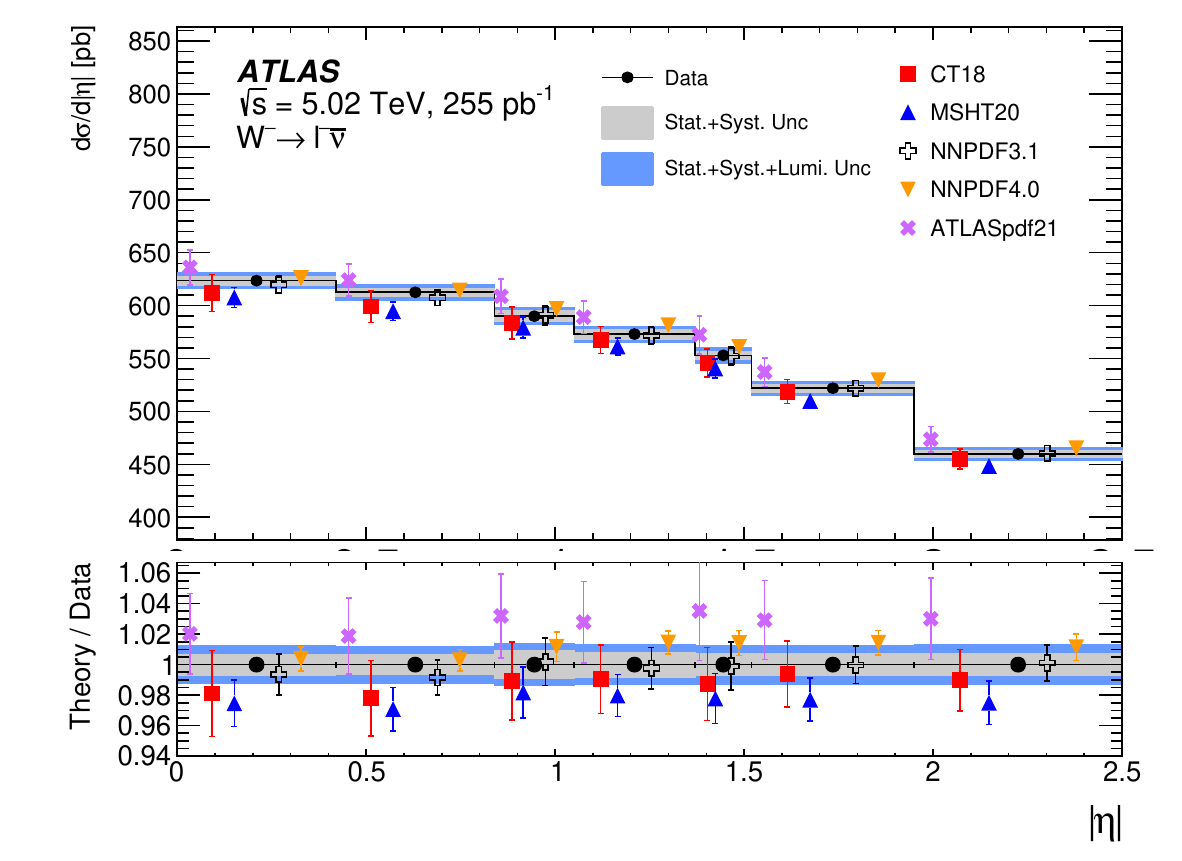}}
\subfloat[]{\includegraphics[width=0.495\linewidth]{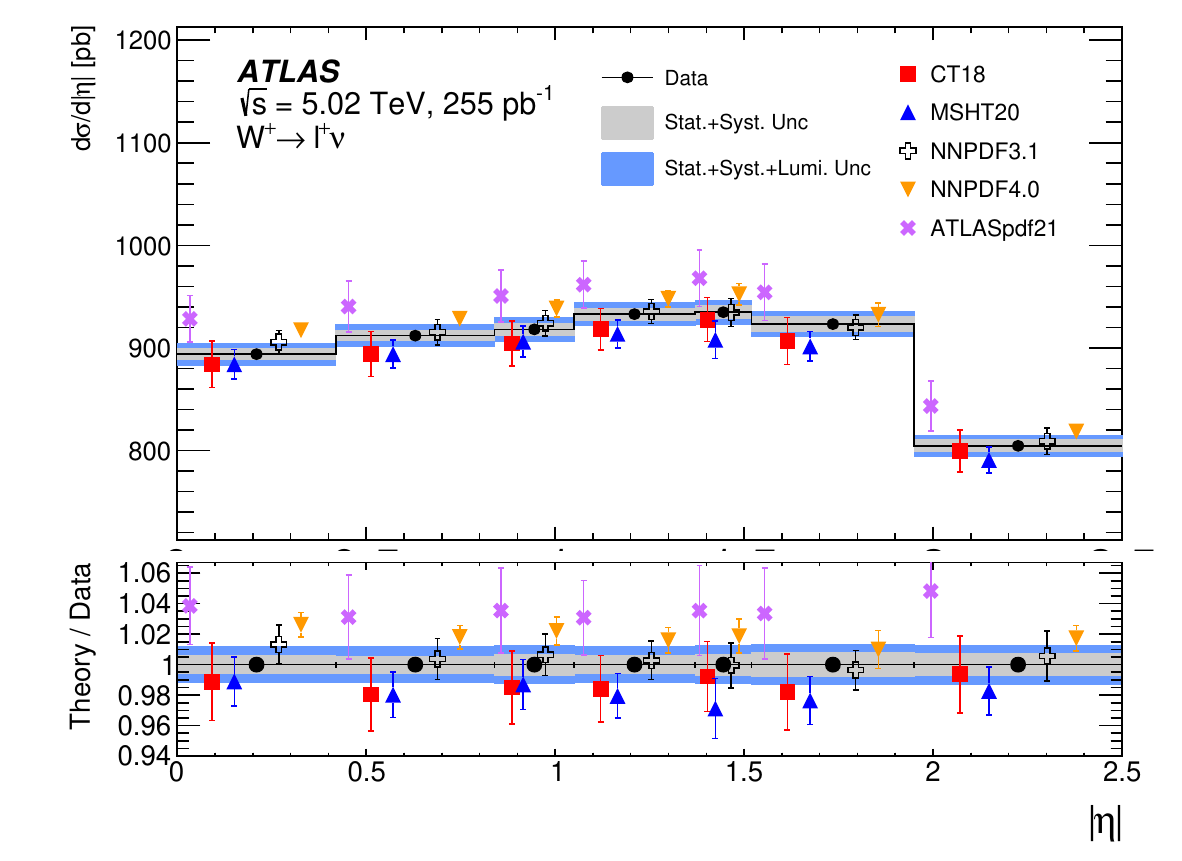}} \\
\subfloat[]{\includegraphics[width=0.495\linewidth]{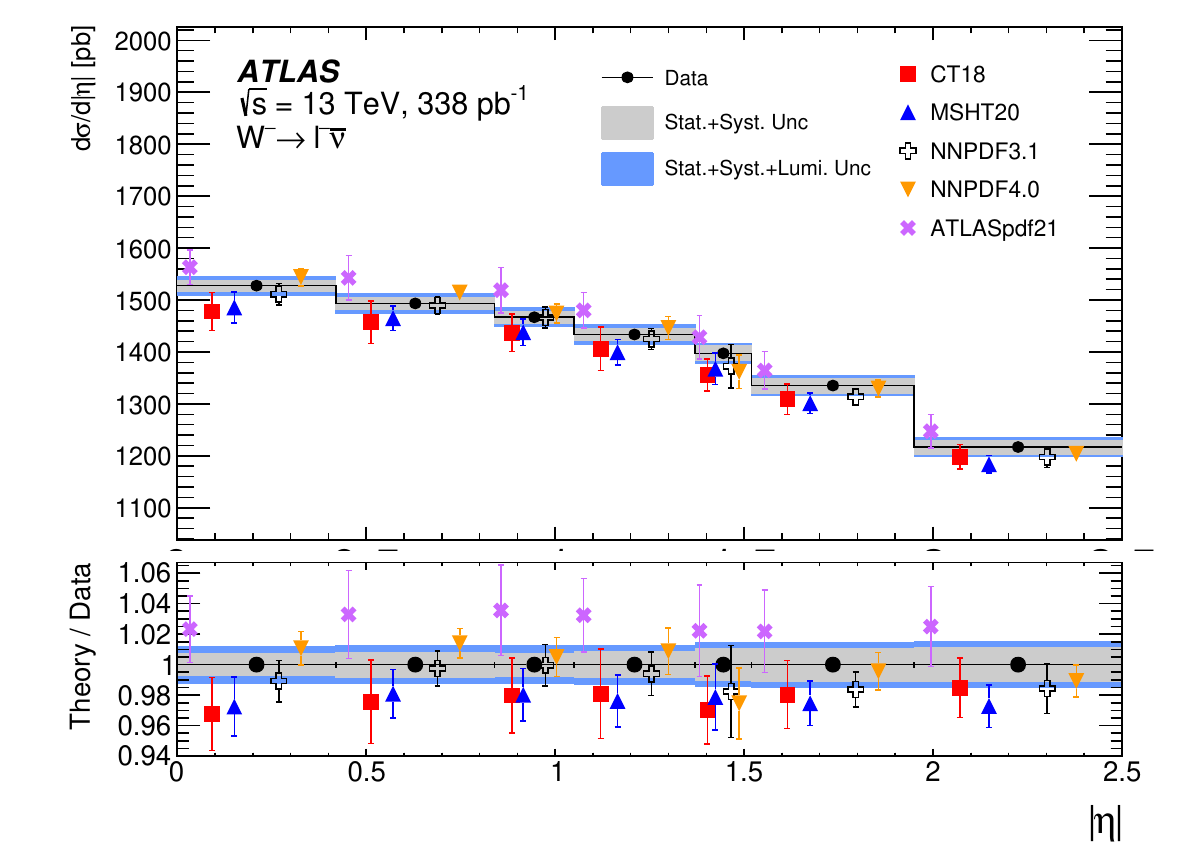}}
\subfloat[]{\includegraphics[width=0.495\linewidth]{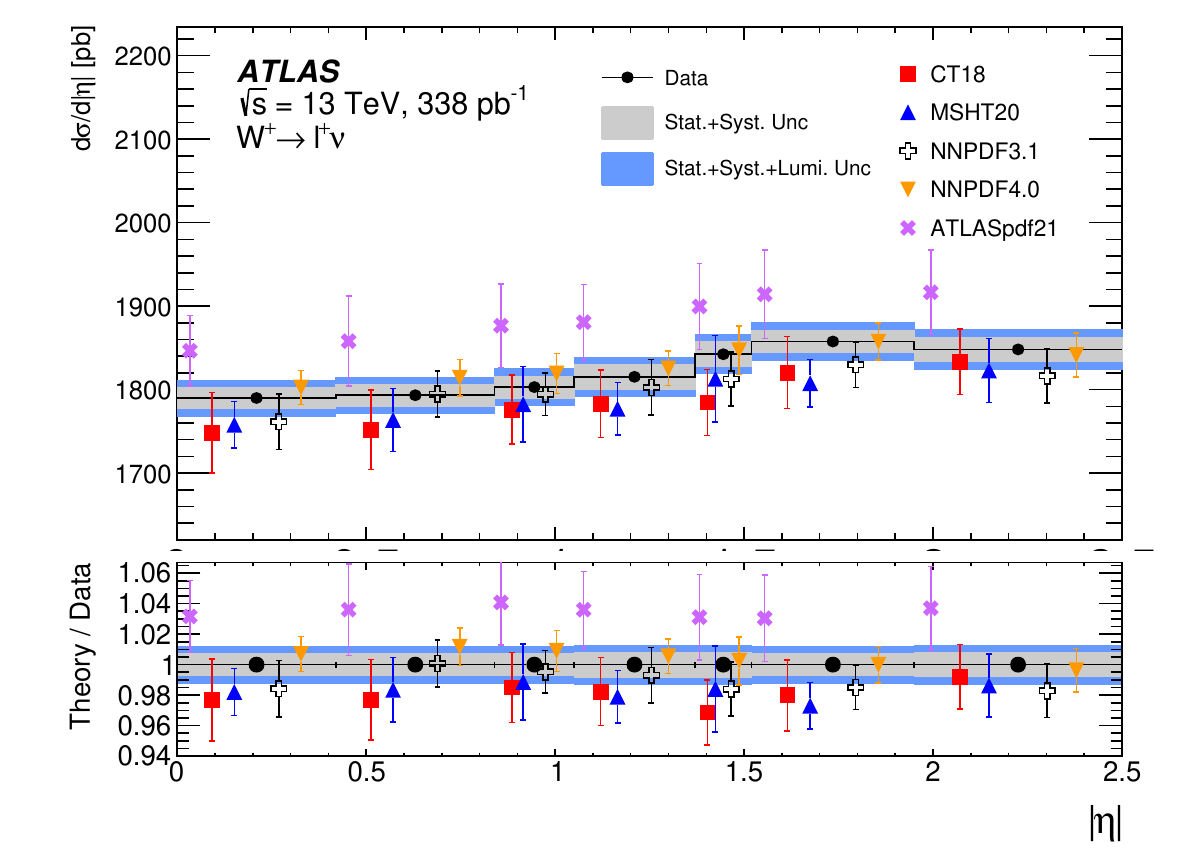}} \\
\caption{Differential (a, c) $W^-$ and (b, d) $W^+$-boson cross-section as a function of $\AbsEta$ for the combination between the electron and muon channels at (a, b) 5.02\,TeV and (c, d) 13\,TeV. The data are presented with systematic (excluding luminosity) and total uncertainties and compared with the theory predictions with the corresponding uncertainties calculated with \textsc{DYTurbo} with different PDF sets.  The ratios of the predictions to the data are shown in the lower panels.}%
\label{fig:diff_result_combined_eta}
\end{figure}

\begin{figure}[tbp]
\centering
\subfloat[]{\includegraphics[width=0.495\linewidth]{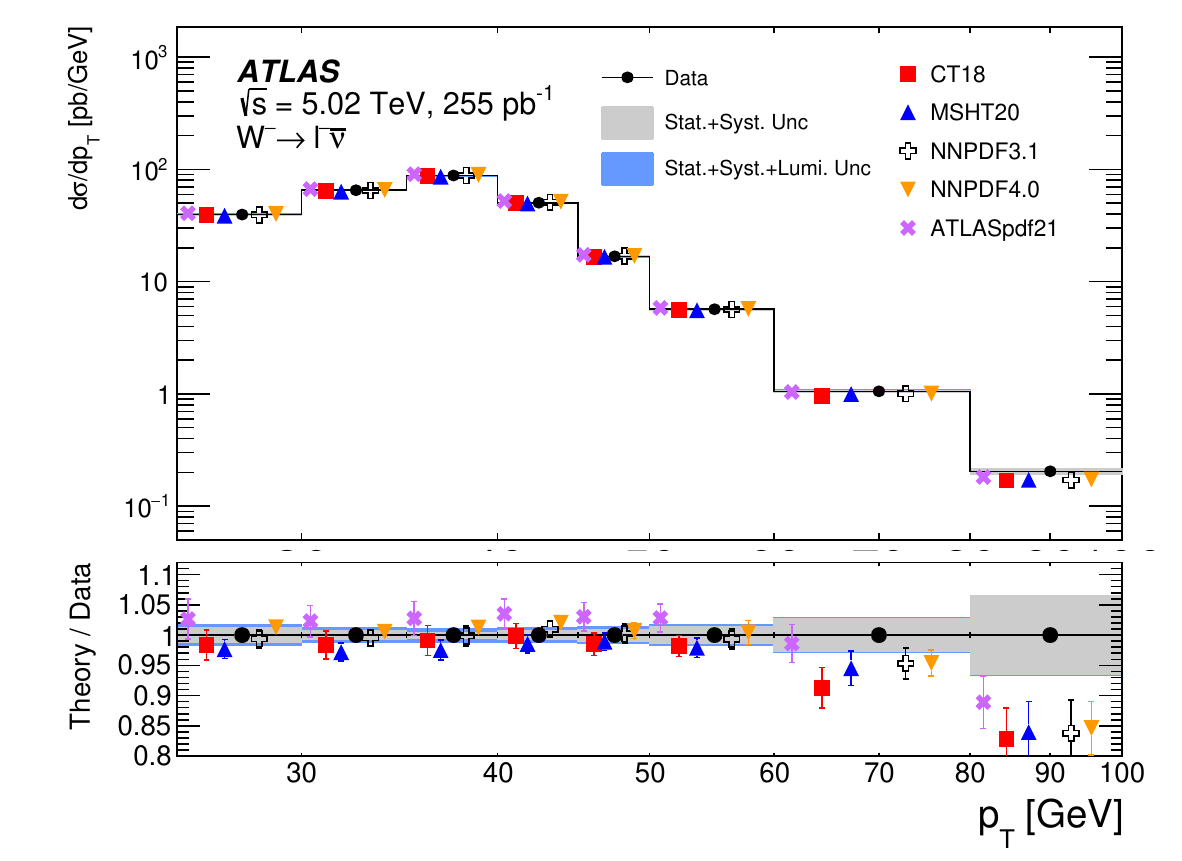}}
\subfloat[]{\includegraphics[width=0.495\linewidth]{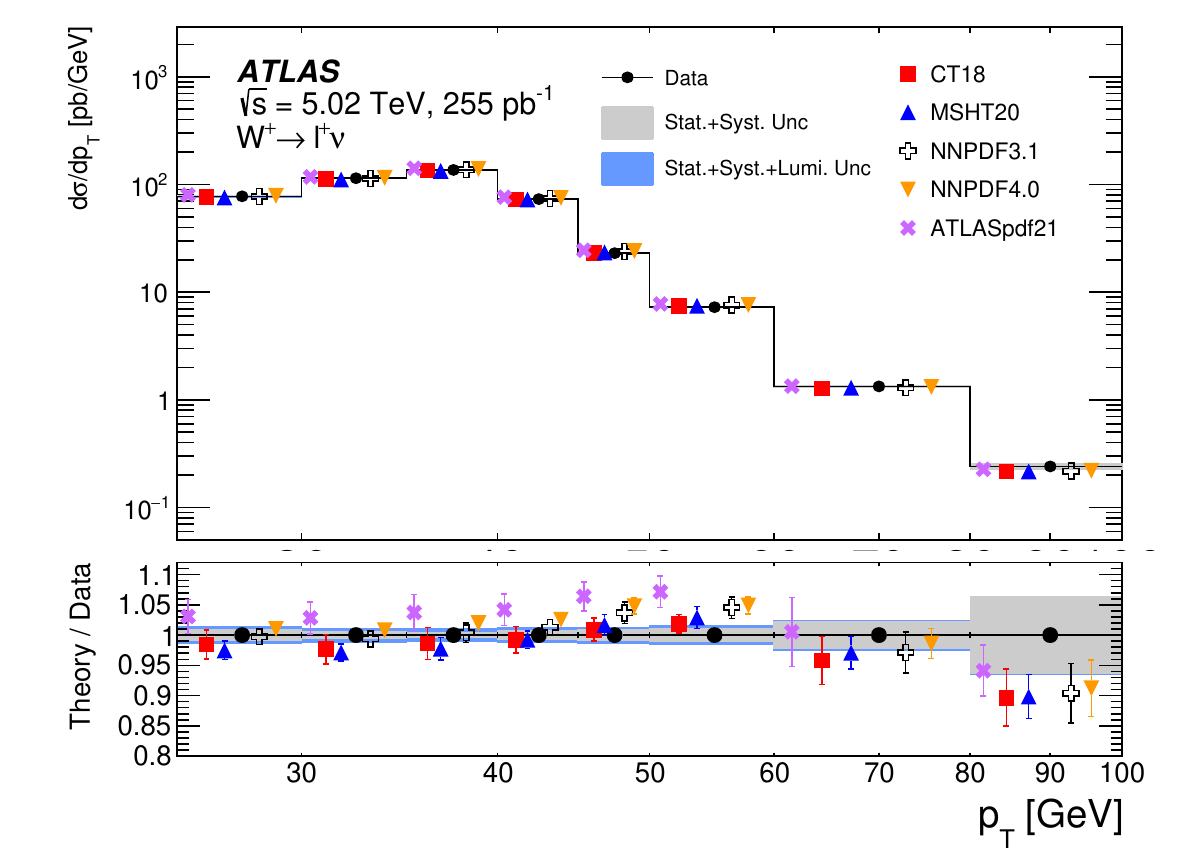}} \\
\subfloat[]{\includegraphics[width=0.495\linewidth]{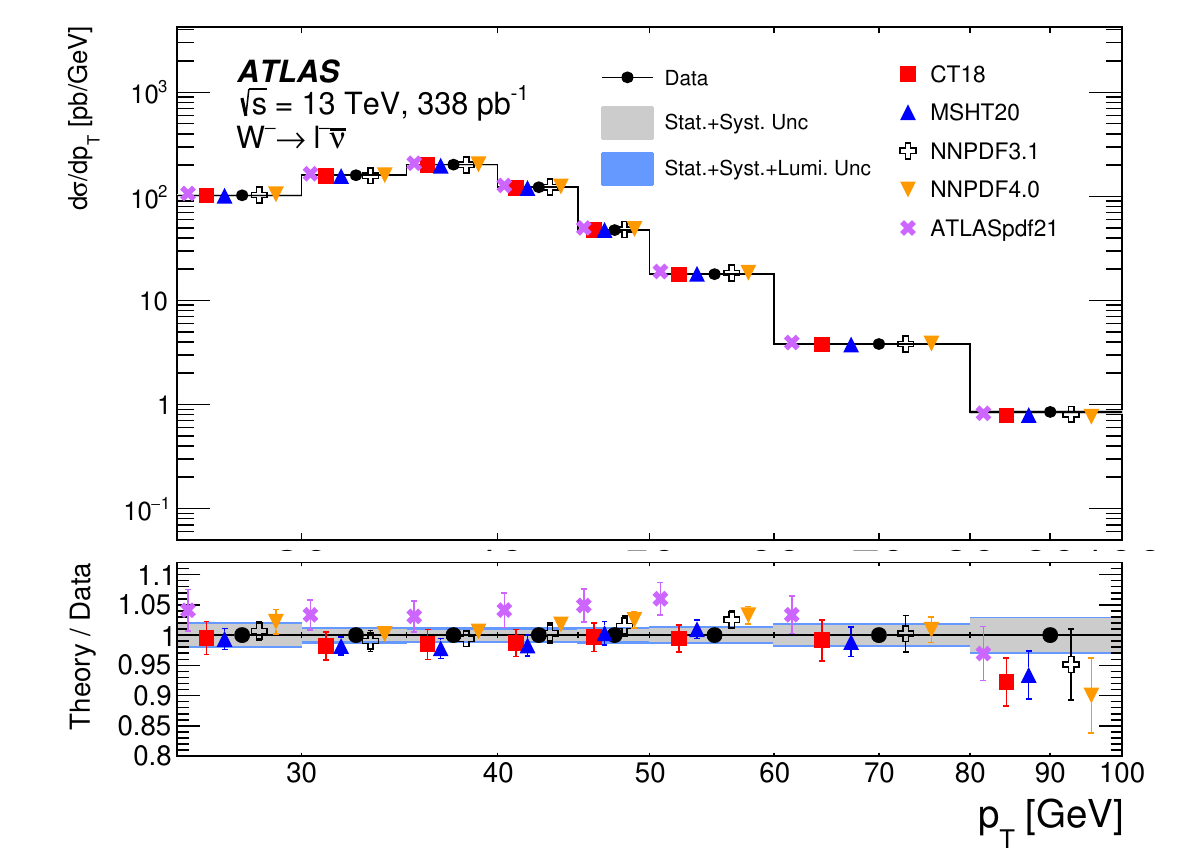}}
\subfloat[]{\includegraphics[width=0.495\linewidth]{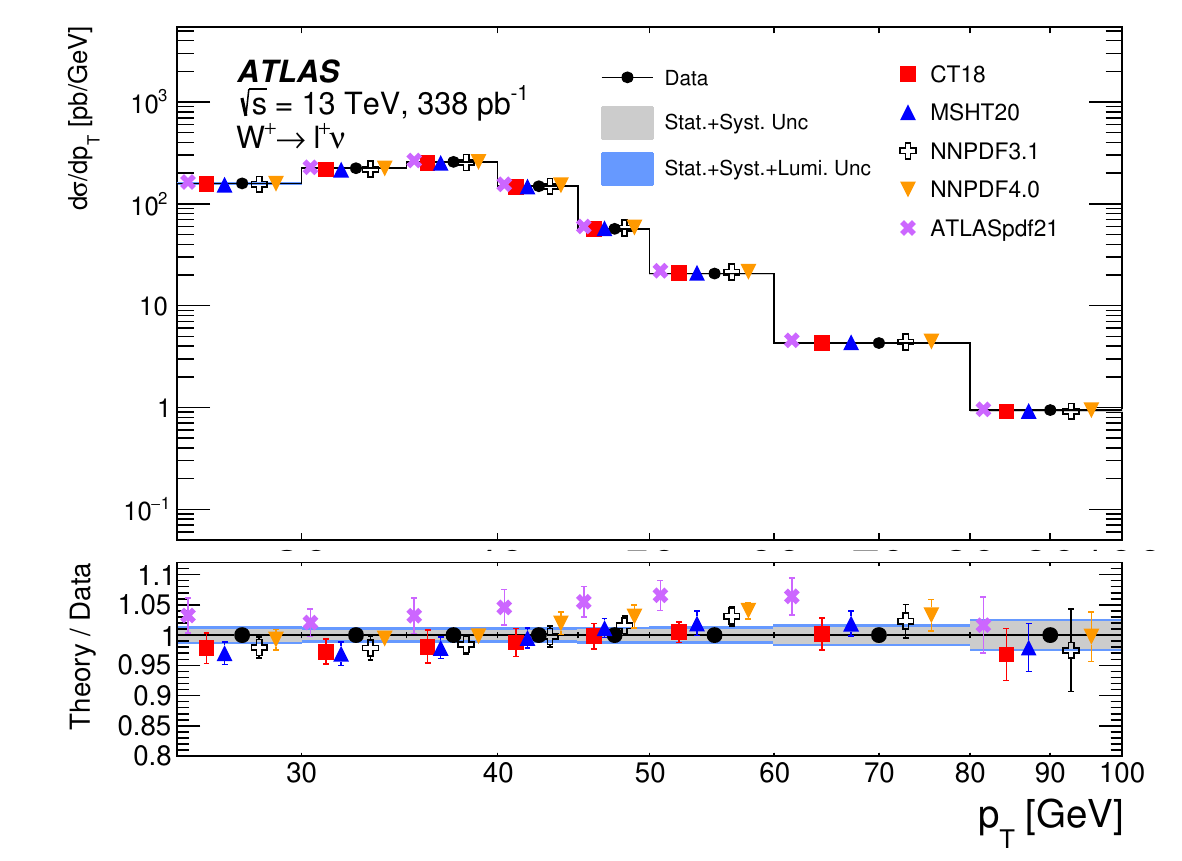}} \\
\caption{Differential (a, c) $W^-$ and (b, d) $W^+$-boson cross-section as a function of $\pTl$ for the combination between the electron and muon channels at (a, b) 5.02\,TeV and (c, d) 13\,TeV. The data are presented with systematic (excluding luminosity) and total uncertainties and compared with the theory predictions with the corresponding uncertainties calculated with \textsc{DYTurbo} with different PDF sets.  The ratios of the predictions to the data are shown in the lower panels.}%
\label{fig:diff_result_combined_ptl}
\end{figure}

\begin{figure}[tbp]
\centering
\subfloat[]{\includegraphics[width=0.95\linewidth]{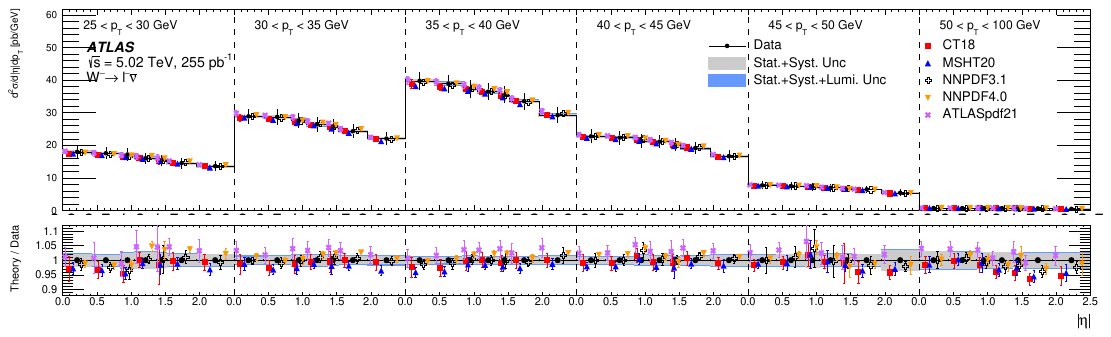}} \\
\subfloat[]{\includegraphics[width=0.95\linewidth]{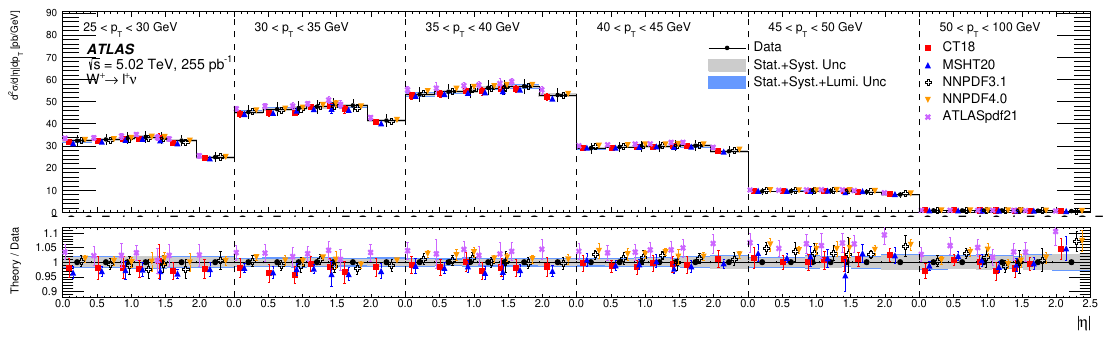}} \\
\subfloat[]{\includegraphics[width=0.95\linewidth]{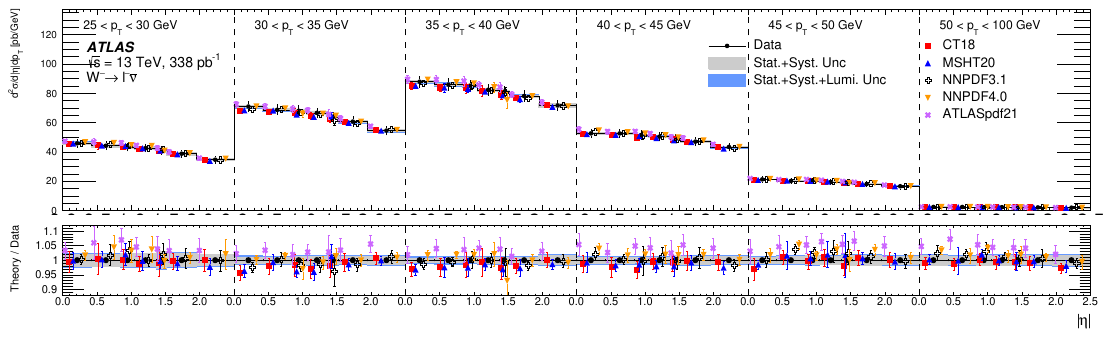}} \\
\subfloat[]{\includegraphics[width=0.95\linewidth]{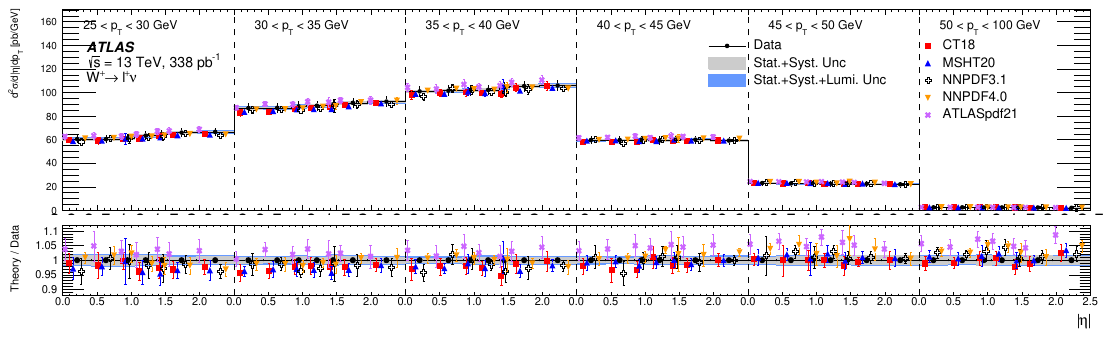}} \\
\caption{Differential (a, c) $W^-$ and (b, d) $W^+$-boson cross-section as functions of $\AbsEta$ and $\pTl$ for the combination between the electron and muon channels at (a, b) 5.02\,TeV and (c, d) 13\,TeV. The data are presented with systematic (excluding luminosity) and total uncertainties and compared with the theory predictions with the corresponding uncertainties calculated with \textsc{DYTurbo} with different PDF sets.  The ratios of the predictions to the data are shown in the lower panels.}%
\label{fig:diff_result_combined_ptl_eta}
\end{figure}

The combined single-differential cross-sections are shown as a function of lepton $|\eta|$ in Figure~\ref{fig:diff_result_combined_eta} and lepton $p_\mathrm{T}$ in Figure~\ref{fig:diff_result_combined_ptl}.\footnote{Unlike the previous ATLAS $W$-boson differential cross-section publications (e.g.\ Ref.~\cite{STDM-2012-20}), the differential cross-sections as a function of lepton $|\eta|$ presented here are the sum of contributions from leptons with negative and positive $\eta$, rather than the average contributions.}
The total uncertainty for the former is around 1.2\% with little variation as a function of lepton $|\eta|$ and is dominated by the luminosity uncertainty, while the latter has a similar precision around the Jacobian peak and increases slightly up to 2\% at low lepton $p_\mathrm{T}$ and up to 6\% for 5.02\,TeV and 3\% for 13\,TeV at the highest lepton $p_\mathrm{T}$ bin, where the dominant uncertainties are from the choice of generator and the hadronic recoil measurement.
The combined two-dimensional cross-sections as functions of $|\eta|$ and $p_\mathrm{T}$ are  presented in Figure~\ref{fig:diff_result_combined_ptl_eta}.
The total uncertainty of the combined double-differential cross-sections is typically around 2\% and is dominated at low $p_\mathrm{T}$ by the lepton-related uncertainty for 5.02\,TeV and the multijet background uncertainty for 13\,TeV and at high $p_\mathrm{T}$ by the data statistical and generator uncertainties.

The measured cross-sections are compared with the theoretical predictions calculated using \textsc{DYTurbo}~\cite{Camarda:2019zyx,Camarda:2021ict,Camarda:2021jsw,Camarda:2023dqn} with recent NNLO PDF sets: CT18~\cite{Hou:2019efy}, MSHT20~\cite{Cridge:2021qfd,Bailey:2020ooq,Cridge:2021pxm}, NNPDF3.1~\cite{NNPDF:2017mvq}, NNPDF4.0~\cite{Ball:2021leu} and ATLASpdf21~\cite{STDM-2020-32}.
The corresponding uncertainties of the predictions include the statistical and systematic ones accounting for variations in the PDF input, $\alpha_\mathrm{s}$, and QCD renormalisation and factorisation scales to cover for missing higher-order contributions.
The calculation precision is NNLO in QCD with NNLL terms in the fiducial phase space.
The prediction with NNPDF4.0 has the highest precision, followed by NNPDF3.1, MSHT20, CT18 and ATLASpdf2.1 due to different input datasets and tolerance values used in the definition of $1\sigma$ error.
The predictions with NNPDF3.1 and NNPDF4.0 provide the best description of the $d\sigma/d|\eta|$ measurements over the full range and of $d\sigma/dp_\mathrm{T}$ for lepton $p_\mathrm{T}$ up to 45\,GeV.
Good agreement is also observed between the measurements and the predictions with the other PDF sets in all the channels and at both energies except in the high $p_\mathrm{T}$ tail where all the predictions are somewhat lower than the measurement in particular for the 5.02\,TeV $W^-$-boson channel.


\subsection{$W$-boson charge asymmetry}%
\label{sub:asymmetry}

\begin{figure}[!htb]
\centering
\subfloat[]{\includegraphics[width=0.495\linewidth]{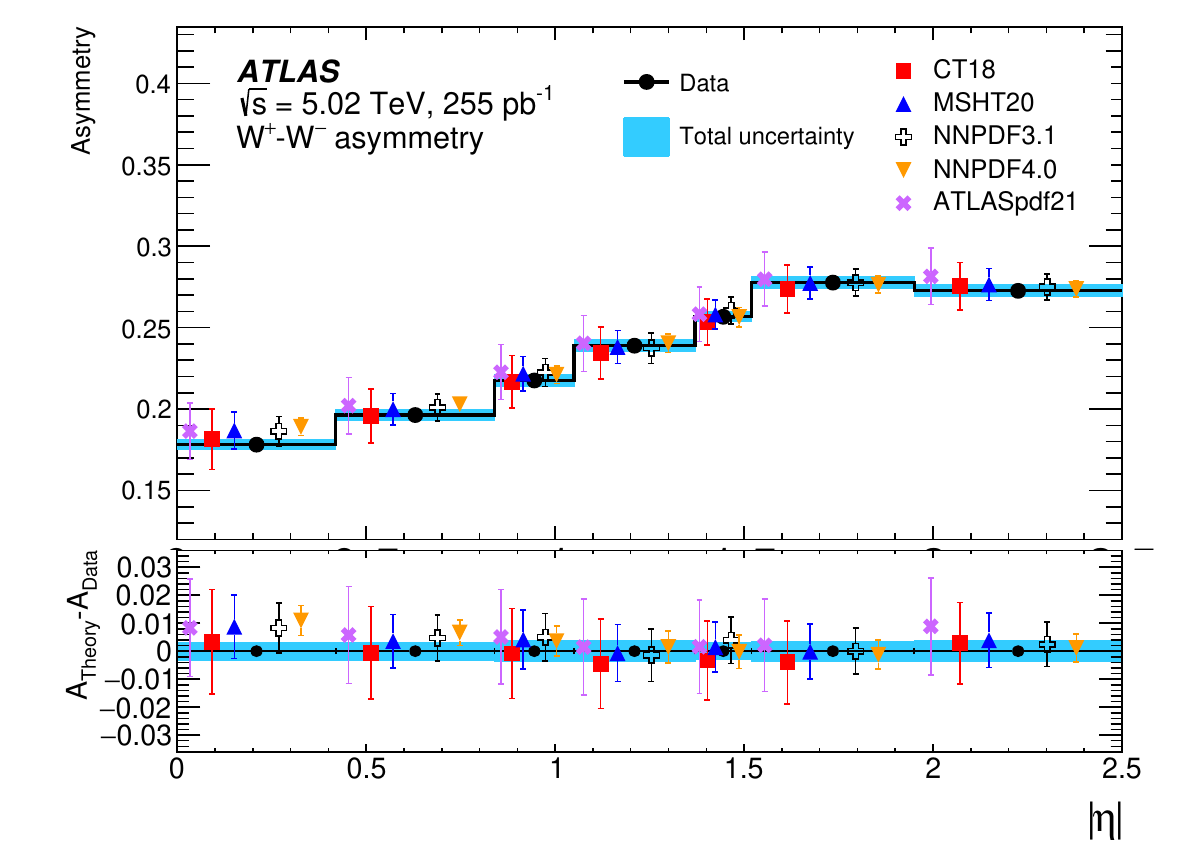}}
\subfloat[]{\includegraphics[width=0.495\linewidth]{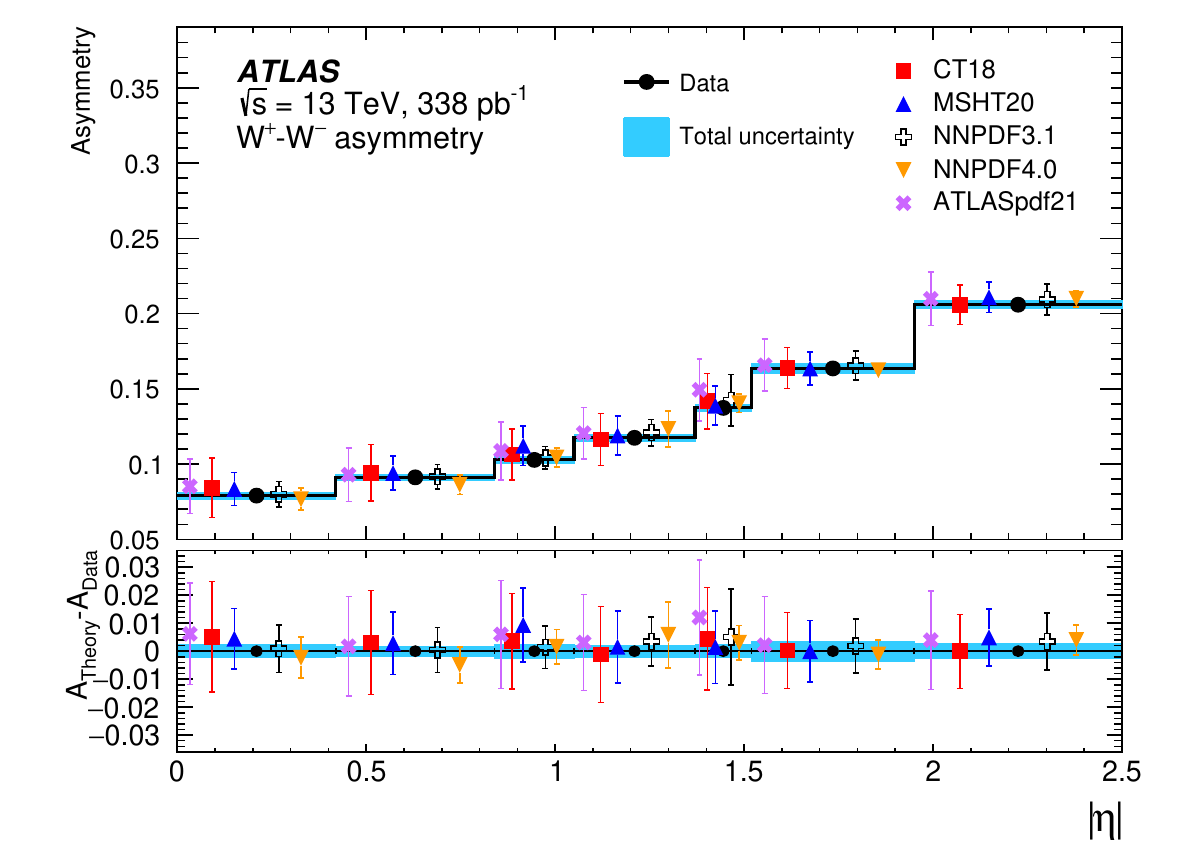}}
\caption{The measured $W$-boson charge asymmetry as a function of $\AbsEta$ for the combined electron and muon channels at (a) 5.02\,TeV and (b) 13\,TeV. The data are presented with total uncertainties and compared with the theory predictions with the corresponding uncertainties using different PDF sets.  The differences between the predictions and the measurement are shown in the lower panels.}%
\label{fig:diff_result_Asymmetry}
\end{figure}

Using Eq.~(\ref{eq:asymmetry}) in Section~\ref{sec:procedure}, the $W$-boson charge asymmetry $A_\ell$ as a function of lepton $|\eta|$ is derived from the corresponding $W^+$ and $W^-$-boson single-differential cross-sections.
The correlations in uncertainties between the two measurements are taken into account, which significantly reduce the experimental uncertainties which are dominated by the statistical and generator  uncertainties. The total uncertainty is below 1.8\% and 3\% for 5.02\,TeV and 13\,TeV, respectively.
Figure~\ref{fig:diff_result_Asymmetry} presents the comparison between measured and predicted $W^\pm$-boson charge asymmetry.
Good agreement is observed in the full $|\eta|$ range and the measurements are more precise than the predictions, demonstrating the potential for additional constraints on the PDFs.

\subsection{PDF profiling}%
\label{sub:PDF profiling}

Before investigating the impact of the cross-sections $d\sigma/d|\eta|$ on PDFs, it is possible to  quantitatively compare the agreement between these cross-sections and the predictions using the recent NNLO PDF sets in terms of $\chi^2$ defined as
\begin{equation}
\chi^2(\vec{\beta}^\mathrm{exp},\vec{\beta}^\mathrm{th})=\sum_{i}^{N_\mathrm{data}}\left(\frac{\sigma_i^\mathrm{exp} - \sigma_i^\mathrm{th}\left(1- \sum_j \beta_j^\mathrm{exp}\gamma_{ij}^\mathrm{exp} -  \sum_k \beta_k^\mathrm{th}\gamma_{ik}^\mathrm{th}\right)}{\Delta_i}\right)^2 + \sum_j^{N_\mathrm{exp,sys}}\left(\beta_j^\mathrm{exp}\right)^2+\sum_k^{N_\mathrm{th,sys}}\left(\beta_k^\mathrm{th}\right)^2\,, \label{eq:chi2}
\end{equation}
where the index $i$ runs over all $N_\mathrm{data}$ data points~\cite{STDM-2012-20}. The measurements and the theory predictions are given by $\sigma_i^\mathrm{exp}$ and $\sigma_i^\mathrm{th}$, respectively. The correlated experimental and theoretical uncertainties are included using the corresponding vectors $\vec{\beta}^\mathrm{exp}$ and $\vec{\beta}^\mathrm{th}$. Their influence on the data and theory predictions is described by the matrices $\gamma_{ij}^\mathrm{exp}$ and $\gamma_{ik}^\mathrm{th}$, where the indices $j$ and $k$ correspond to the experimental ($N_\mathrm{exp,sys}$) and theoretical ($N_\mathrm{th,sys}$) uncertainty sources.
The statistical and uncorrelated experimental uncertainties are included in $\Delta_i$.
For the experimental uncertainties, the correlation model described in Section~\ref{sec:Systematic uncertainties} is employed.

The PDF sets considered are CT18, MSHT20, NNPDF4.0 and ATLASpdf21. These PDF sets are determined using datasets that include the LHC data from $W$-boson production at $\sqrt{s}=7$ and $8$\,TeV but not at $\sqrt{s}=5.02$\,TeV and $13$\,TeV.
The theoretical prediction is calculated using \textsc{MadGraph5\_aMC@NLO}~\cite{Alwall:2014hca} at NLO level, which is interfaced to \textsc{APPLgrid}~\cite{Carli:2010rw} using \textsc{aMCfast}~\cite{Bertone:2014zva}. In a further step, the bin-by-bin NNLO corrections are applied by comparing NLO predictions with NNLO+NNLO predictions calculated using \textsc{DYTurbo} for the same PDF, in terms of $K$-factors, $K^\mathrm{NLO\to\ NNLO\ QCD}$:
\begin{equation}
K_i^\mathrm{NLO\to NNLO\ QCD} = (d\sigma^\mathrm{NNLO+NNLL\ QCD}_\mathrm{PDF}/d|\eta|)_i / (d\sigma^\mathrm{NLO\ QCD}_\mathrm{PDF}/d|\eta|)_i\,,
\end{equation}
where $i$ denotes the bin number.
The resulting $\chi^2$ values per number of degrees of freedom (ndof) comparing the measured cross-sections and predictions for each of the PDF sets are shown in Table~\ref{tab:pdf_profiling_chi2}.
The partial $\chi^2$ value for each dataset corresponds to that of the first term on the right-hand side of  Eq.~(\ref{eq:chi2}), while the correlated $\chi^2$ value is from the other two terms.
The measurements are well described by these PDFs in particular for CT18 and MSHT20 given their large tolerance values used in the uncertainty definition, while they exhibit slight tension with ATLASpdf21 and NNPDF4.0.

The impact of new data on these PDF sets can be estimated in a quantitative way with a profiling technique~\cite{Paukkunen:2014zia,STDM-2012-20}. The profiling is performed in the \textsc{xFitter}  framework~\cite{H1:2009bcq,H1:2018flt,alekhin2015herafitter} using the above $\chi^2$ formula, except that the correlated experimental and theory uncertainties are now treated as nuisance parameters. When the $\chi^2$ function is minimised, the values of the nuisance parameters at the minimum can be interpreted as an optimisation (`profiling') of PDFs to describe the data~\cite{Paukkunen:2014zia}.

\begin{figure}[tbp]
\centering
\subfloat[]{\includegraphics[width=0.33\linewidth]{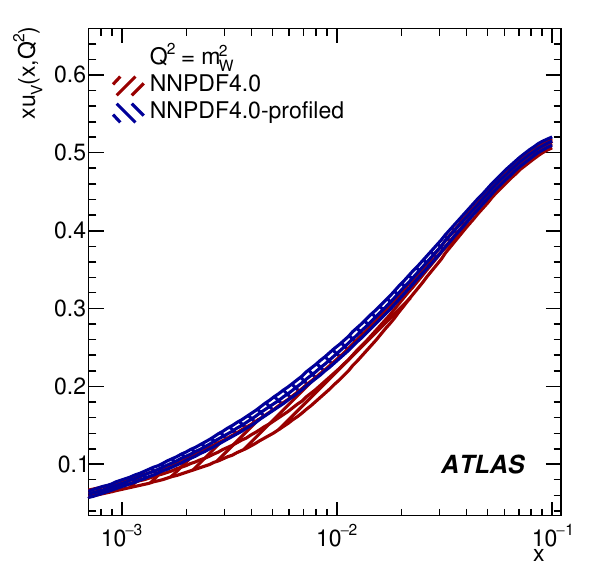}}
\subfloat[]{\includegraphics[width=0.33\linewidth]{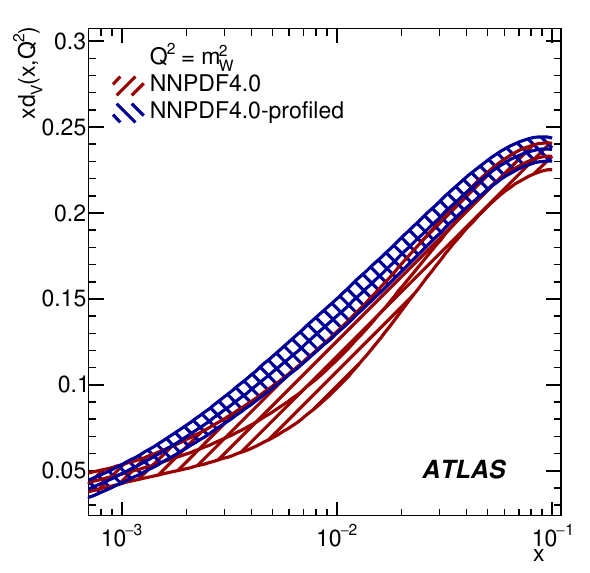}}
\subfloat[]{\includegraphics[width=0.33\linewidth]{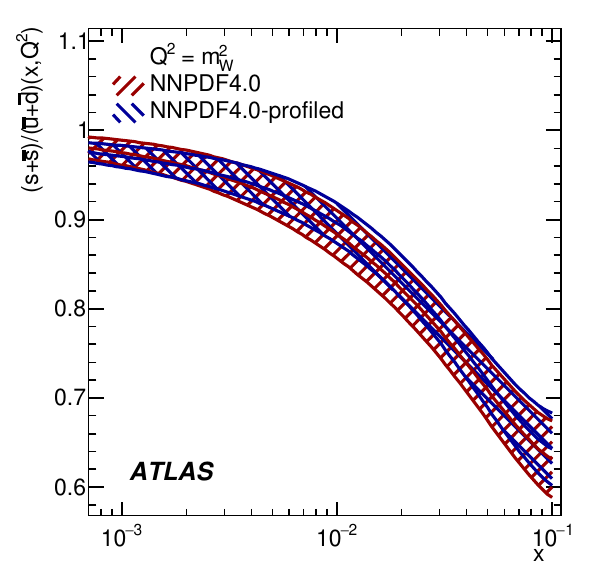}}
\caption{Original (red) and profiled (blue) PDF distributions of the NNPDF4.0 PDF set at $Q^2=m_W^2$ using the measured differential cross-sections in $\AbsEta$ at 5.02\,TeV and 13\,TeV. The distributions of (a) $xu_v$, (b) $xd_v$, and (c) $(s+\bar{s})/(\bar{u}+\bar{d})$ PDFs are presented.}
\label{fig:profiler_NNPDF40_q2_6464}
\end{figure}

\begin{figure}[tbp]
\centering
\subfloat[]{\includegraphics[width=0.33\linewidth]{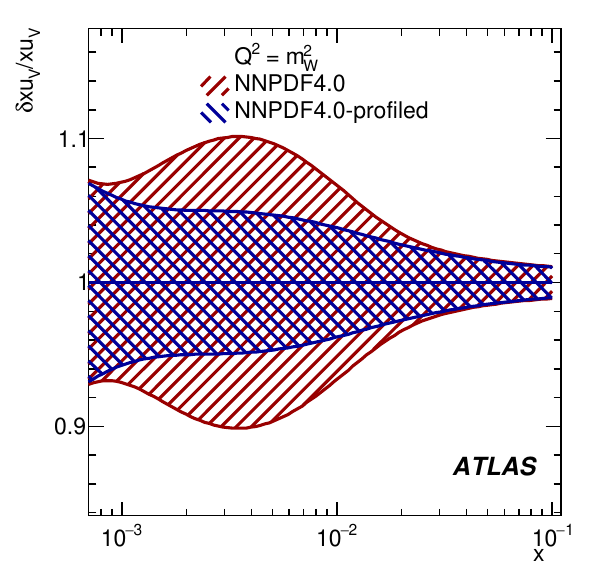}}
\subfloat[]{\includegraphics[width=0.33\linewidth]{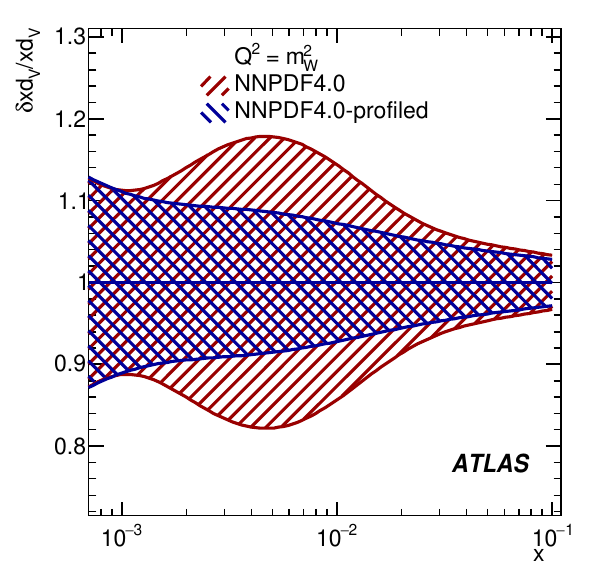}}
\subfloat[]{\includegraphics[width=0.33\linewidth]{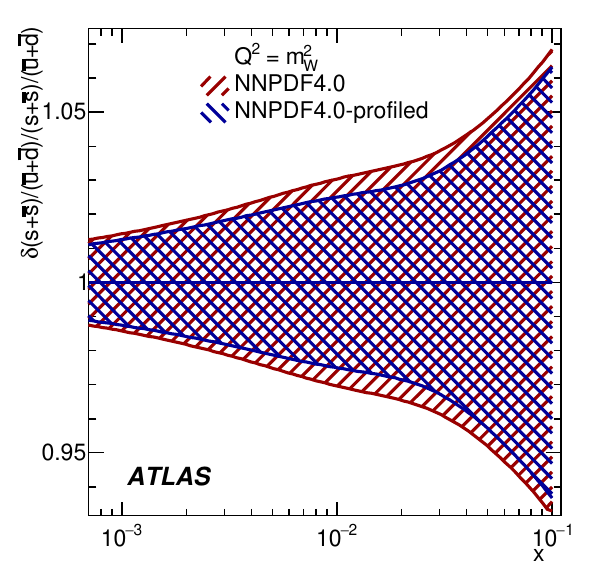}}
\caption{The relative uncertainties of original (red) and profiled (blue) PDF distributions of the NNPDF4.0 PDF set at $Q^2=m_\mathrm{W}^2$ using the measured differential cross-sections in $\AbsEta$ at 5.02\,TeV and 13\,TeV for (a) $xu_v$, (b) $xd_v$, and (c) $(s+\bar{s})/(\bar{u}+\bar{d})$.}
\label{fig:profiler_NNPDF40_q2_6464_ratio}
\end{figure}

Taking NNPDF4.0 as an example, the profiled parton distributions with reduced uncertainties incorporating the new measurements in comparison with the original ones are shown in Figure~\ref{fig:profiler_NNPDF40_q2_6464}
for up- and down-valence quarks ($xu_v$ and $xd_v$) as well as the ratio of strange quark to light sea quark distributions ($R_s=(s+\bar{s})/(\bar{u}+\bar{d})$).
The latter was chosen since it has been observed~\cite{STDM-2012-14,STDM-2019-18} to be higher in some PDF fits than previously expected.
Figure~\ref{fig:profiler_NNPDF40_q2_6464_ratio} displays the corresponding change on the relative uncertainties of the parton distributions.
Most of the constraining power arises from the $\sqrt{s} = 13$\,TeV measurement. Sizeable improvements in the uncertainties in the quark distributions are observed.
The largest reduction of about a factor two is achieved around $x=0.004$ for the valence quarks.
Similar improvement is also observed for the CT18, MSHT20 and ATLASpdf21 PDF sets.

\begin{table}[tbp]
\centering
\caption{Values of $\chi^2$ with the corresponding numbers of degree of freedom (ndof) comparing the data and unprofiled predictions with various PDF sets. The $\chi^2$ values are split by dataset, providing a detailed breakdown of the agreement between the data and predictions. Additionally, the contribution of the correlated terms arising from the experimental and theoretical correlated uncertainties is also listed.
}
\label{tab:pdf_profiling_chi2}

\begin{tabular}{c|c|c|c|c|c}
\hline
Dataset            & Ndof & CT18      & MSHT20    & ATLASpdf21 & NNPDF4.0  \\
\hline
\Wmtolnu\ 5.02 TeV  & 7     & 1.7 & 2.0  & 6.0   & 6.7   \\
\Wptolnu\ 5.02 TeV  & 7     & 3.7 & 5.1  & 6.8    & 6.1   \\
\Wmtolnu\ 13 TeV    & 7     & 7.0  & 7.6  & 7.5   & 9.2   \\
\Wptolnu\ 13 TeV    & 7     & 4.0 & 3.9  & 5.6   & 3.5  \\
\hline
Correlated $\chi^2$ &       & 11   & 16    & 30    & 25    \\
\hline
Total $\chi^2$      & 28    & 28   & 35   & 56     & 50    \\
\hline
\end{tabular}


\end{table}


%

\clearpage
\section{Conclusions}
\label{sec:conclusion}

High precision single-differential cross-sections as a function of lepton transverse momentum  $p_\mathrm{T}$ or lepton pseudorapidity $|\eta|$ in the electron and muon channels of the $W^\pm$-boson decays, as well as double-differential cross-sections of these variables, are measured in proton--proton collisions at $\sqrt{s}=5.02$\,TeV and 13\,TeV.
The single-differential cross-sections in $p_\mathrm{T}$ and the double-differential cross-sections are the first such measurements from ATLAS.
They are based on data samples collected in dedicated low pile-up runs with the ATLAS detector at the LHC, corresponding to integrated luminosities of 255\,pb$^{-1}$ and 338\,pb$^{-1}$, respectively. The cross-sections are measured to an experimental precision better than 0.5\% in the bulk phase space, apart from the normalisation uncertainty of the luminosity determination, which is 1\% and 0.92\% for the 5.02\,TeV and 13\,TeV datasets, respectively.
The combined cross-sections of the electron and muon channels are compared with predictions at NNLO  including transverse momentum resummation at NNLL accuracy in QCD using different PDF sets and overall good agreement is observed.
The $W$-boson charge asymmetry as a function of lepton $|\eta|$ is also measured at both centre-of-mass energies and the measurements are in good agreement with the corresponding predictions.
In addition, the expected impact of the measured single-differential cross-sections as a function of lepton $|\eta|$ on several recent PDF sets is estimated, showing their constraining power on various quark PDFs.


%

%
%
%

%
%

%
%
%
%

%

%
%
%

%
%
%
%

%

%
%
%
%
%
\section*{Acknowledgements}
%

%

%
%

%
%

We thank CERN for the very successful operation of the LHC and its injectors, as well as the support staff at
CERN and at our institutions worldwide without whom ATLAS could not be operated efficiently.

The crucial computing support from all WLCG partners is acknowledged gratefully, in particular from CERN, the ATLAS Tier-1 facilities at TRIUMF/SFU (Canada), NDGF (Denmark, Norway, Sweden), CC-IN2P3 (France), KIT/GridKA (Germany), INFN-CNAF (Italy), NL-T1 (Netherlands), PIC (Spain), RAL (UK) and BNL (USA), the Tier-2 facilities worldwide and large non-WLCG resource providers. Major contributors of computing resources are listed in Ref.~\cite{ATL-SOFT-PUB-2025-001}.

We gratefully acknowledge the support of ANPCyT, Argentina; YerPhI, Armenia; ARC, Australia; BMWFW and FWF, Austria; ANAS, Azerbaijan; CNPq and FAPESP, Brazil; NSERC, NRC and CFI, Canada; CERN; ANID, Chile; CAS, MOST and NSFC, China; Minciencias, Colombia; MEYS CR, Czech Republic; DNRF and DNSRC, Denmark; IN2P3-CNRS and CEA-DRF/IRFU, France; SRNSFG, Georgia; BMBF, HGF and MPG, Germany; GSRI, Greece; RGC and Hong Kong SAR, China; ICHEP and Academy of Sciences and Humanities, Israel; INFN, Italy; MEXT and JSPS, Japan; CNRST, Morocco; NWO, Netherlands; RCN, Norway; MNiSW, Poland; FCT, Portugal; MNE/IFA, Romania; MSTDI, Serbia; MSSR, Slovakia; ARIS and MVZI, Slovenia; DSI/NRF, South Africa; MICIU/AEI, Spain; SRC and Wallenberg Foundation, Sweden; SERI, SNSF and Cantons of Bern and Geneva, Switzerland; NSTC, Taipei; TENMAK, T\"urkiye; STFC/UKRI, United Kingdom; DOE and NSF, United States of America.

Individual groups and members have received support from BCKDF, CANARIE, CRC and DRAC, Canada; CERN-CZ, FORTE and PRIMUS, Czech Republic; COST, ERC, ERDF, Horizon 2020, ICSC-NextGenerationEU and Marie Sk{\l}odowska-Curie Actions, European Union; Investissements d'Avenir Labex, Investissements d'Avenir Idex and ANR, France; DFG and AvH Foundation, Germany; Herakleitos, Thales and Aristeia programmes co-financed by EU-ESF and the Greek NSRF, Greece; BSF-NSF and MINERVA, Israel; NCN and NAWA, Poland; La Caixa Banking Foundation, CERCA Programme Generalitat de Catalunya and PROMETEO and GenT Programmes Generalitat Valenciana, Spain; G\"{o}ran Gustafssons Stiftelse, Sweden; The Royal Society and Leverhulme Trust, United Kingdom.

In addition, individual members wish to acknowledge support from Armenia: Yerevan Physics Institute (FAPERJ); CERN: European Organization for Nuclear Research (CERN DOCT); Chile: Agencia Nacional de Investigaci\'on y Desarrollo (FONDECYT 1230812, FONDECYT 1230987, FONDECYT 1240864); China: Chinese Ministry of Science and Technology (MOST-2023YFA1605700, MOST-2023YFA1609300), National Natural Science Foundation of China (NSFC - 12175119, NSFC 12275265, NSFC-12075060); Czech Republic: Czech Science Foundation (GACR - 24-11373S), Ministry of Education Youth and Sports (ERC-CZ-LL2327, FORTE CZ.02.01.01/00/22\_008/0004632), PRIMUS Research Programme (PRIMUS/21/SCI/017); EU: H2020 European Research Council (ERC - 101002463); European Union: European Research Council (ERC - 948254, ERC 101089007, ERC, BARD, 101116429), Horizon 2020 Framework Programme (MUCCA - CHIST-ERA-19-XAI-00), European Union, Future Artificial Intelligence Research (FAIR-NextGenerationEU PE00000013), Italian Center for High Performance Computing, Big Data and Quantum Computing (ICSC, NextGenerationEU); France: Agence Nationale de la Recherche (ANR-20-CE31-0013, ANR-21-CE31-0013, ANR-21-CE31-0022, ANR-22-EDIR-0002); Germany: Baden-Württemberg Stiftung (BW Stiftung-Postdoc Eliteprogramme), Deutsche Forschungsgemeinschaft (DFG - 469666862, DFG - CR 312/5-2); Italy: Istituto Nazionale di Fisica Nucleare (ICSC, NextGenerationEU), Ministero dell'Università e della Ricerca (NextGenEU PRIN20223N7F8K M4C2.1.1); Japan: Japan Society for the Promotion of Science (JSPS KAKENHI JP22H01227, JSPS KAKENHI JP22H04944, JSPS KAKENHI JP22KK0227, JSPS KAKENHI JP23KK0245); Norway: Research Council of Norway (RCN-314472); Poland: Ministry of Science and Higher Education (IDUB AGH, POB8, D4 no 9722), Polish National Science Centre (NCN 2021/42/E/ST2/00350, NCN OPUS 2023/51/B/ST2/02507, NCN OPUS nr 2022/47/B/ST2/03059, NCN UMO-2019/34/E/ST2/00393, UMO-2020/37/B/ST2/01043, UMO-2022/47/O/ST2/00148, UMO-2023/49/B/ST2/04085, UMO-2023/51/B/ST2/00920); Portugal: Foundation for Science and Technology (FCT); Spain: Generalitat Valenciana (Artemisa, FEDER, IDIFEDER/2018/048), Ministry of Science and Innovation (MCIN \& NextGenEU PCI2022-135018-2, MICIN \& FEDER PID2021-125273NB, RYC2019-028510-I, RYC2020-030254-I, RYC2021-031273-I, RYC2022-038164-I); Sweden: Carl Trygger Foundation (Carl Trygger Foundation CTS 22:2312), Swedish Research Council (Swedish Research Council 2023-04654, VR 2018-00482, VR 2021-03651, VR 2022-03845, VR 2022-04683, VR 2023-03403), Knut and Alice Wallenberg Foundation (KAW 2018.0458, KAW 2019.0447, KAW 2022.0358); Switzerland: Swiss National Science Foundation (SNSF - PCEFP2\_194658); United Kingdom: Leverhulme Trust (Leverhulme Trust RPG-2020-004), Royal Society (NIF-R1-231091); United States of America: U.S. Department of Energy (ECA DE-AC02-76SF00515), Neubauer Family Foundation.

%
%


%
%
%
%
%
%

%

%
%
%
\printbibliography
\clearpage
 
\begin{flushleft}
\hypersetup{urlcolor=black}
{\Large The ATLAS Collaboration}

\bigskip

\AtlasOrcid[0000-0002-6665-4934]{G.~Aad}$^\textrm{\scriptsize 104}$,
\AtlasOrcid[0000-0001-7616-1554]{E.~Aakvaag}$^\textrm{\scriptsize 17}$,
\AtlasOrcid[0000-0002-5888-2734]{B.~Abbott}$^\textrm{\scriptsize 123}$,
\AtlasOrcid[0000-0002-0287-5869]{S.~Abdelhameed}$^\textrm{\scriptsize 119a}$,
\AtlasOrcid[0000-0002-1002-1652]{K.~Abeling}$^\textrm{\scriptsize 55}$,
\AtlasOrcid[0000-0001-5763-2760]{N.J.~Abicht}$^\textrm{\scriptsize 49}$,
\AtlasOrcid[0000-0002-8496-9294]{S.H.~Abidi}$^\textrm{\scriptsize 30}$,
\AtlasOrcid[0009-0003-6578-220X]{M.~Aboelela}$^\textrm{\scriptsize 45}$,
\AtlasOrcid[0000-0002-9987-2292]{A.~Aboulhorma}$^\textrm{\scriptsize 36e}$,
\AtlasOrcid[0000-0001-5329-6640]{H.~Abramowicz}$^\textrm{\scriptsize 157}$
% ...
\bigskip
\\

% ...
$^{17}$Department for Physics and Technology, University of Bergen, Bergen; Norway.\\
% ...
$^{30}$Physics Department, Brookhaven National Laboratory, Upton NY; United States of America.\\
% ...
$^{36}$$^{(a)}$Facult\'e des Sciences Ain Chock, Universit\'e Hassan II de Casablanca;$^{(b)}$Facult\'{e} des Sciences, Universit\'{e} Ibn-Tofail, K\'{e}nitra;$^{(c)}$Facult\'e des Sciences Semlalia, Universit\'e Cadi Ayyad, LPHEA-Marrakech;$^{(d)}$LPMR, Facult\'e des Sciences, Universit\'e Mohamed Premier, Oujda;$^{(e)}$Facult\'e des sciences, Universit\'e Mohammed V, Rabat;$^{(f)}$Institute of Applied Physics, Mohammed VI Polytechnic University, Ben Guerir; Morocco.\\
% ...
$^{45}$Physics Department, Southern Methodist University, Dallas TX; United States of America.\\
% ...
$^{49}$Fakult\"{a}t Physik , Technische Universit{\"a}t Dortmund, Dortmund; Germany.\\
% ...
$^{55}$II. Physikalisches Institut, Georg-August-Universit\"{a}t G\"ottingen, G\"ottingen; Germany.\\
% ...
$^{104}$CPPM, Aix-Marseille Universit\'e, CNRS/IN2P3, Marseille; France.\\
% ...
$^{119}$$^{(a)}$New York University Abu Dhabi, Abu Dhabi;$^{(b)}$United Arab Emirates University, Al Ain; United Arab Emirates.\\
% ...
$^{123}$Homer L. Dodge Department of Physics and Astronomy, University of Oklahoma, Norman OK; United States of America.\\
% ...
$^{157}$Raymond and Beverly Sackler School of Physics and Astronomy, Tel Aviv University, Tel Aviv; Israel.\\
% ...
$^{a}$ Also at Affiliated with an institute formerly covered by a cooperation agreement with CERN.\\
$^{b}$ Also at An-Najah National University, Nablus; Palestine.\\
$^{c}$ Also at Borough of Manhattan Community College, City University of New York, New York NY; United States of America.\\
$^{d}$ Also at Center for High Energy Physics, Peking University; China.\\
$^{e}$ Also at Center for Interdisciplinary Research and Innovation (CIRI-AUTH), Thessaloniki; Greece.\\
$^{f}$ Also at Centre of Physics of the Universities of Minho and Porto (CF-UM-UP); Portugal.\\
% ...

\end{flushleft}


%

%
%
%
%

%
%
%
%

\end{document}